\documentclass[aps,prd, reprint, showpacs,superscriptaddress,
floatfix,amsmath,amssymb]{revtex4-1}
\usepackage{lipsum}
\usepackage{multirow}
\usepackage{graphicx}
\usepackage{caption}
\usepackage{subcaption}
\usepackage{bm}
\usepackage{color}

\input epsf

\usepackage{hyperref}

\begin{document}

\date{\today}

\title{Testing scalar-tensor theories and PPN parameters in Earth orbit}

\author{Andreas Sch\"arer}
\email{andreas.schaerer@physik.uzh.ch}
\affiliation{Department of Physics, University of Zurich,
Winterthurerstrasse 190, 8057 Zurich, Switzerland}

\author{Raymond Ang\'elil}
\affiliation{Institute for Computational Science, University of Zurich,
Winterthurerstrasse 190, 8057 Zurich, Switzerland}

\author{Ruxandra Bondarescu}
\affiliation{Department of Physics, University of Zurich,
Winterthurerstrasse 190, 8057 Zurich, Switzerland}

\author{Philippe Jetzer}
\affiliation{Department of Physics, University of Zurich,
Winterthurerstrasse 190, 8057 Zurich, Switzerland}

\author{Andrew Lundgren}
\affiliation{Max Planck Institute for Gravitational Physics 
(Albert Einstein Institute), Callinstr. 38, 30167 Hannover, Germany}

\begin{abstract}
We compute the PPN parameters $\gamma$ and $\beta$ for general scalar-tensor 
theories in the Einstein frame, which we compare to the existing
PPN formulation in the Jordan frame for alternative 
theories of gravity. 
This computation is important for scalar-tensor theories that are 
expressed in the Einstein frame, such as chameleon and symmetron theories,
which can incorporate hiding mechanisms that predict 
environment-dependent PPN parameters.
We introduce a general formalism for scalar-tensor theories
and constrain it using the limit on $\gamma$ given by the
Cassini experiment.
In particular we discuss massive Brans-Dicke scalar fields for
extended sources.
Next, using a recently proposed Earth satellite experiment, 
in which atomic clocks are used for spacecraft tracking, we compute the 
observable perturbations in the redshift induced by PPN parameters deviating 
from their general relativistic values.
Our estimates suggest that 
$|\gamma - 1| \sim |\beta -1|  \sim 10^{-6}$ may be detectable by a
satellite that carries a clock with fractional frequency uncertainty 
$\Delta f/f \sim 10^{-16}$ in an eccentric orbit around the Earth. 
Such space experiments are within reach of existing atomic clock technology. 
We discuss further the requirements necessary for such a mission to detect 
deviations from Einstein relativity.

\end{abstract}

\maketitle

\section{Introduction}
General relativity (GR) is the widely accepted theory to explain gravitation.
Nonetheless, there are other theories of gravity which also satisfy 
the experimental constraints and remain candidates for the correct theory of 
gravity. These theories are being constrained by various high precision
experiments. In particular, the recent development of ultra-precise frequency 
standards and atom interferometers provide new opportunities for testing
different aspects of gravity. In this paper, we focus on scalar-tensor 
theories, which are a class of alternative theories of gravity that in addition
to the metric tensor include a scalar field. 
We are interested both in developing 
the theoretical framework for testing these theories and in
estimating potential constraints from upcoming satellite missions that carry 
clocks in space. 

Scalar-tensor theories are widely used in particle physics, string theory and 
cosmology to model poorly understood phenomena
for which we may have some observations such as in the case of dark matter and 
dark energy, but the new physics remains tantalisingly
just out of reach. Effective scalar fields can arise from underlying, not-yet 
understood fundamental physics such as 
compactified extra dimensions \cite{compactified.extra.dimensions} or string 
theory, which includes the dilaton scalar field \cite{Damour}. 
Since the detection of the Higgs particle \cite{higgs} we know that scalar 
(spin-0) particles exist in nature.
The phase of inflation \cite{inflation.guth}, a short period of 
rapid expansion in the very early universe, could have been
caused by a scalar field. 
Quintessence models make use of scalar fields 
causing the late-time acceleration of the 
universe and therefore they could replace the cosmological 
constant and explain dark energy \cite{Quintessence.and.the.Rest.of.the.World}.
These scalar fields may couple to matter in ways
that slightly violate general relativity and could be detected as our 
instrumentation becomes more precise.

The most accessible testbed for theories of gravity is the external environment
of compact bodies such as binary pulsars or solar system objects.
Here the gravitational field is weak, allowing the use of the 
parametrized post-Newtonian formalism (PPN).
While there are 
infinitely many possible frames, typically, 
the Lagrangians of these theories are expressed either in the Jordan or the 
Einstein frame. In the Jordan frame, the scalar field multiplies 
the Ricci scalar and any present matter fields couple directly to the frame 
metric, while in the Einstein frame the Ricci
scalar appears alone (as in traditional Einstein gravity) and the matter 
fields couple to a conformally related metric.  We focus
on the $\gamma$ and $\beta$ parameters predicted by scalar-tensor theories 
for which we have existent experimental constraints.

The simplest scalar-tensor theory is the original Brans-Dicke theory, 
where the massless scalar field and its constant coupling function (in the
Jordan frame) lead to 
$\gamma= (\omega_0+1)/(\omega_0+2)$ and $\beta=1$ \cite{will.review}. 
For the case of a massive Brans-Dicke field, which contains a potential 
$U\sim m^2 \varphi^2$ in addition to the constant 
coupling, these parameters were determined in \cite{periv}. 
The introduction of the mass 
term induces the $\gamma$-parameter to become distance dependent.
The parameters for chameleon theories were derived in \cite{hees}. In 
special cases, 
PPN parameters have been calculated for more general theories such as 
scalar-tensor-vector theories where, as its name implies, an additional 
vector field enters the stage of gravitation \cite{vector}.
For general scalar-tensor theories formulated in the Jordan frame, they
were determined in Hohmann {\it et al.} \cite{hohmann.ppn}. 

We calculate the PPN parameters for a general scalar-tensor theory
expressed in the Einstein frame.
We show how this formalism can be
useful for finding the PPN parameters for specific choices of 
scalar-tensor theories. 
Next, turning 
to a variety of scalar-tensor theories which predict constant PPN parameters, 
we investigate the near-future prospects for the measurement 
of deviations from Einstein relativity using a class of Earth-orbiting atomic 
clock experiments introduced in our earlier work, Ang\'elil {\it et al.}
\cite{clock.paper}.

The PPN parameters are typically calculated for a spacetime consisting of a 
point source surrounded by vacuum. This assumption is, in general, not 
appropriate to solve the scalar field equation of motion. For example in 
chameleon scalar field theories \cite{chameleon.khoury.prl,*chameleon.khoury} 
or symmetron theories \cite{symmetron.prl,*symmetron.cosmology}, the field 
behaves in a complex way inside the Earth due to its high density, 
significantly altering the external
field profile. This also implies that the PPN parameters can depend on the 
environment. Performing an experiment around Earth may reveal different PPN 
parameters compared to the same experiment performed in the vicinity 
of the Sun.

Therefore, to discuss constraints on the PPN parameters in general we
introduce a simple formalism containing a free parameter which 
can depend on the properties of both the theory and the source mass
under consideration.
It can account for the effects arising from the finite size of the source
like screening effects in chameleon theories.
We solve the scalar field equation for a massive Brans-Dicke scalar field
in- and outside a sphere of constant density and show that our ansatz
indeed represents the typical field profile of a massive scalar.
The most stringent constraint in the solar system comes from 
measurements of the Cassini spacecraft, which limit the 
size of the $\gamma$ PPN parameter around the Sun \cite{cassini.nature}.
We use this limit to improve constraints on massive Brans-Dicke 
theory discussed in \cite{periv,hohmann.ppn} by regarding 
the Sun as a homogeneous sphere instead of a point source.
However, while the Sun has low density, it is not an ideal candidate to probe 
theories that propose hiding mechanisms due to its high compactness $M/R$.
The Earth and the Moon are more suitable due to their lower compactness to 
test theories such as chameleon theories. 
 
In the second part of the paper, we bring attention to the increasing accuracy 
of space-qualified atomic clocks. Our estimates show that a
space clock that can reach the accuracy of the Atomic Clock Ensemble in Space
\cite{ACES} $\Delta f/f \sim 10^{-16}$ in an eccentric orbit around the Earth 
could place constraints on the $\beta$ and $\gamma$ PPN parameters 
around the Earth of about $10^{-6}$ over the course of one orbit.
It can be expected that in future, many space missions will use either an 
ultra-precise atomic clock or a transponder that can reflect signals from 
other clocks on Earth and in space to track the spacecraft.
These will allow the ability to constrain or detect signals from alternative 
theories of gravity. The estimates presented 
here are obtained by taking the difference between the redshift signal 
generated by general relativity $\gamma = \beta =1$, and the signal 
generated by a theory with $\gamma$ and $\beta$ different from one. The 
numbers obtained here are large enough to suggest detectability if 
a clock-carrying mission on an orbit like that of the originally proposed 
satellite Space-Time Explorer and QUantum Equivalence Principle Test 
(STE-QUEST) were to fly \cite{STE-QUEST.Yellow.book,*SQ.science}. 
However, to make any definitive 
statements further work that aims to recover the signal of specific alternative
theory of gravity from realistic data would be needed. 
We show that the difference in the redshift signal between general relativity 
and a small deviation peaks around pericenter. We study the width of these
peaks to find the time scale which needs to be resolved in order to be 
sensitive to such deviations. 
 
The outline of this paper is as follows. In section 
\ref{sec:Parametrized Post-Newtonian Formalism}, the parametrized 
post-Newtonian formalism is briefly reviewed.
Section \ref{sec:Scalar-Tensor Theory in the Jordan and Einstein Frame} 
discusses the action and the equations of motion of a scalar-tensor theory
in both the Jordan and the Einstein frame. The conformal transformation 
relating these frames is addressed, whereas more details can be found in 
appendix \ref{app Conformal Transformation between Jordan and Einstein Frame}. 
After briefly reviewing the procedure to obtain the PPN parameters in the 
Jordan frame in \ref{sec:PPN Parameter from the Jordan Frame}, we calculate 
these parameters in detail for any theory formulated in the Einstein frame in 
section \ref{sec:PPN Parameter from the Einstein Frame}.
In section \ref{Framework to Constrain PPN Parameters}, we address constraints 
on scalar theories. We bring attention to the importance of screening 
mechanisms and propose a simple framework to constrain scalar theories and 
discuss current and future experimental constraints. Note that in this paper 
we restrict attention to local constraints on the PPN parameters and do not 
discuss cosmological constraints on chameleon models.  We apply this formalism 
to some examples: (A) Brans-Dicke theory, the simplest case of a scalar-tensor 
theory, (B) massless fields with a more general coupling, (C) massive
Brans-Dicke theory and (D) chameleon 
fields, an example for a field with a screening mechanism.
Next, in section \ref{Constraining PPN Parameters in Earth Orbit}, we address 
the possibility of constraining PPN parameters in Earth orbit using satellites 
endowed with state-of-the-art atomic clocks. To do so, we estimate the 
relativistic effects coming from varying PPN parameters using a numerical 
orbit simulation. 

Throughout this work we set the units to $c=\hbar=1$, and 
therefore the reduced Planck mass is $M_\text{Pl} = \sqrt{1/8 \pi G}$.


\section{Scalar-tensor theories in the Einstein frame}
\subsection{The Parametrized Post-Newtonian Formalism}
\label{sec:Parametrized Post-Newtonian Formalism}

The most common way to parametrize theories of gravity in the weak field is 
to use the parametrized post-Newtonian (PPN) formalism \cite{will.book}. 
There, the standard general relativistic metric is generalized with a 
collection of parameters which are permitted to take any value decreed 
by the alternative theory under consideration.

We start with the Schwarzschild metric written in isotropic coordinates
$(t,\chi,\theta,\varphi)$
\begin{align}
\begin{split}
ds^2 &= g_{\mu\nu} dx^\mu dx^\nu
\\
&= - \frac{\left(1-\frac{G M}{2\chi}\right)^2}{\left(1+\frac{G M}{2\chi}
\right)^2}dt^2
+ {\left(1+\frac{G M}{2\chi}\right)^4} (d\chi^2 + \chi^2 d\Omega^2),
\end{split}
\end{align}
where $d\Omega := d\theta^2+\sin^2\theta d\varphi^2$.
This is the vacuum solution of the Einstein field equations outside a
spherically symmetric non-charged and non-rotating mass $M$.
$G$ denotes the Newtonian gravitational constant.
In this paper, we are interested in solar system constraints and since within
the solar system gravitational fields are weak and typical velocities are 
small, it is sufficient to consider the post-Newtonian limit of this metric.
To do so, we introduce a parameter $\epsilon$.
Its power tracks the order of a term, where 
$\epsilon \sim G M/r$, although numerically $\epsilon = 1$.
Massive particles moving on an orbit typically have velocities 
$v^2 \approx G M/r$ and therefore $\epsilon \sim v^{2}$ (Note that other 
authors use the convention $\epsilon \sim v$).
Typically, we have $G M/r \ll 1$ within the solar system.
Therefore, after endowing the different terms in the Schwarzschild metric 
with the appropriate $\epsilon^n$, we can perform an expansion in 
$\epsilon$ and neglect higher order terms.
For the post-Newtonian level, we keep terms up to order $\epsilon^2$ in 
$g_{00}$ and up to order $\epsilon$ in $g_{ij}$.
Many alternative theories of gravity predict solutions which start to deviate
from the ones predicted by general relativity at this level.
Therefore, the parameters $\gamma$ and $\beta$ are added to the metric
to model deviations from general relativity ($\gamma = \beta = 1$)
\cite{will.book}.
Here, we promote the $\gamma$ and $\beta$ from constants to functions
of $\chi$. Additionally, the gravitational `constant' is allowed to change with
distance.

This gives the metric
\begin{align}
\label{equ Jordan frame metric ansatz}
\begin{split}
ds_\text{J}^2 = &- \left( 1 - h_{\text{J}00}^{(1)}(\chi) \epsilon
- h_{\text{J}00}^{(2)}(\chi) \epsilon^2 \right) dt_\text{J}^2
\\&+ \left( 1 + h_{\text{J}\chi\chi}^{(1)}(\chi) \epsilon \right) 
\left( d\chi^2 + \chi^2 d\Omega^2 \right)
\end{split}
\end{align}
with
\begin{align}
\begin{split}
\label{equ Jordan frame metric components}
h_{\text{J}00}^{(1)}(\chi) &= \frac{2 G_\text{J}(\chi) M_\text{J}}{\chi}
\\
h_{\text{J}\chi\chi}^{(1)}(\chi) &= \gamma(\chi)\frac{2 G_\text{J}(\chi)
M_\text{J}}{\chi}
\\
h_{\text{J}00}^{(2)}(\chi) &= -\beta(\chi) \frac{4 G_\text{J}^2(\chi) 
M^2_\text{J}}{2\chi^2}.
\end{split}
\end{align}
The index $J$ indicates that this metric is formulated in the 
Jordan frame (see next section).
Note that if more intricate spacetimes are considered, additional
parameters may enter the metric.



\subsection{The choice of frame}
\label{sec:Scalar-Tensor Theory in the Jordan and Einstein Frame}

The action of a scalar-tensor theory can be written in various ways. In the 
Jordan frame it is
\begin{align}
\begin{split}
\label{STT action J frame}
S = &\int d^4x \sqrt{-g_\text{J}} \frac{M_\text{Pl}^2}{2}
\left[ \varphi R_\text{J} - \frac{\omega(\varphi)}{\varphi} 
\left(\nabla_\text{J} \varphi\right)^2 - U(\varphi) \right]
\\
&+ \int d^{4}x \sqrt{-g_\text{J}} \mathcal{L}_\text{m}^\text{J} 
(\Phi_\text{m},g_{\mu\nu}^\text{J}),
\end{split}
\end{align}
where the theory is characterized by the coupling function $\omega(\varphi)$ 
and the scalar potential $U(\varphi)$, both functions of the scalar field.
The scalar field is considered to be positive everywhere and we assume that 
$U \ge 0$ and $\omega > -3/2$.

There are two characteristic properties of this frame.
First, the non-minimally coupling term $\varphi R_\text{J}$ represents the 
coupling between the scalar field and curvature.
Second, matter fields $\Phi_\text{m}$ couple to the frame metric 
$g_{\mu\nu}^\text{J}$ which is used to determine the Christoffel symbols, 
the Ricci tensor, and to raise and lower indices.
By varying this action with respect to the metric and the scalar field, 
the tensor and the scalar equations of motion
\begin{subequations}
\label{equ:STT EOMs J frame}
\begin{align}
\begin{split}
R_{\mu\nu}^\text{J}
&= \frac{1}{\varphi} \bigg[ 8 \pi G \left( T_{\mu\nu}^\text{J}
- \frac{\omega+1}{2\omega+3} g_{\mu\nu}^\text{J} T_\text{J} \right)
+ \nabla_\mu^\text{J} \partial_\nu \varphi
\\
& + \frac{\omega}{\varphi} \partial_\mu \varphi \partial_\nu \varphi
 \quad- \frac{1}{2} g_{\mu\nu}^\text{J} \frac{1}{2 \omega + 3}  
 \frac{\partial \omega}{\partial \varphi} (\nabla_\text{J} \varphi)^2
\\
&+ \frac{1}{2} g_{\mu\nu}^\text{J} \frac{2 \omega + 1}{2 \omega + 3} U
 + \frac{1}{2} g_{\mu\nu}^\text{J} \frac{1}{2 \omega + 3} \varphi 
 \frac{\partial U}{\partial \varphi} \bigg]
\end{split}
\\
\nabla_\text{J}^2 \varphi
&= \frac{1}{2 \omega + 3} \left( 8\pi G T_\text{J}
- \omega_{,\varphi} \left(\nabla_\text{J}\varphi\right)^2
- 2 U + \varphi U_{,\varphi} \right)
\end{align}
\end{subequations}
are obtained, where 
$\nabla_\text{J}^2 := g^{\mu\nu}_\text{J} \nabla_\mu^\text{J} \partial_\nu$.
By $\nabla_\mu^\text{J}$ we denote the covariant derivative obtained from the 
Jordan frame metric.
By $T_{\mu\nu}^\text{J}$ and $T_\text{J} = g^{\mu\nu}_\text{J} 
T_{\mu\nu}^\text{J}$ we denote the stress-energy tensor and its 
trace in the Jordan frame.

Alternatively, a scalar-tensor theory can be expressed in the Einstein-frame
\begin{align}
\label{STT action E frame}
\begin{split}
S = &\int d^4x \sqrt{-g_\text{E}} \frac{M_\text{Pl}^2}{2} \bigg[ R_\text{E}
 - 2 (\nabla_\text{E} \phi )^2
- V(\phi) \bigg]
\\
&+ \int d^{4}x \sqrt{-g_\text{E}} \mathcal{L}_\text{m}^\text{E} 
(\Phi_\text{m}, F(\phi)^{-1} g_{\mu\nu}^\text{E})
\end{split}
\end{align}
with the corresponding equations of motion
\begin{subequations}
\label{equ:STT EOMs E frame}
\begin{align}
\label{equ:STT EOMs E frame tensor}
R_{\mu\nu}^\text{E}
&= 8 \pi G \left( T_{\mu\nu}^\text{E}
- \frac{1}{2} g_{\mu\nu}^\text{E} T_\text{E} \right)
+ 2 \partial_\mu \phi \partial_\nu \phi
+ \frac{1}{2} g_{\mu\nu}^\text{E} V(\phi)
\\
\nabla_\text{E}^2 \phi
&= \frac{8\pi G}{4} \frac{F_{,\phi}}{F} T_\text{E}
+ \frac{1}{4} V_{,\phi}.
\end{align}
\end{subequations}
Here, the theory is determined by the coupling function $F(\phi)$ and the 
potential $V(\phi)$. In this frame, the field couples minimally to gravity 
and therefore, the gravity part of the action takes the form of the 
Einstein-Hilbert action in general relativity. This comes at the price that 
the matter fields do not couple to the Einstein frame metric directly but 
to the combination $F(\phi)^{-1} g_{\mu\nu}^\text{E}$, and therefore, the 
coupling explicitly depends on the scalar field. But there is an obvious 
advantage when working in the Einstein frame: there, the equations of motion 
\eqref{equ:STT EOMs E frame} are much simpler compared to the ones in the 
Jordan frame \eqref{equ:STT EOMs J frame}, even though these two frames 
are mathematically equivalent.

To avoid confusion between these two frames we label quantities with indices 
J and E, depending on the frame they are coming from.
The two frames are related to each other by a conformal transformation
\begin{align}
\label{equ:conformal transformation}
g_{\mu\nu}^\text{J} = F(\phi)^{-1} g_{\mu\nu}^\text{E}
\end{align}
with $\varphi = F > 0$, i.e. the scalar field in the Jordan frame mimics 
the coupling function in the Einstein frame. The positiveness of the fields 
is required to avoid a change of sign in the metric line element when going 
from one to the other frame. This conformal transformation is discussed in 
appendix \ref{app Conformal Transformation between Jordan and Einstein Frame}.
				

\subsection{PPN parameters in the Jordan frame}
\label{sec:PPN Parameter from the Jordan Frame}

The PPN parameters $\gamma$ and $\beta$ have been calculated for a general 
scalar-tensor theory stated in the Jordan frame \cite{hohmann.ppn}. Here we 
give a very short overview of their derivation. One starts with the ansatz 
for the metric \eqref{equ Jordan frame metric ansatz} where $\chi$ is the 
radial coordinate in isotropic coordinates. Expanding the scalar field 
$\varphi$, the coupling function $\omega$ and the potential $U$ in powers 
of $\epsilon$ (see appendix 
\ref{app Conformal Transformation between Jordan and Einstein Frame}) and 
solving the equations of motion \eqref{equ:STT EOMs J frame} order by order, 
one finds \cite{hohmann.ppn}
\begin{widetext}
\begin{subequations}
\label{equ:PPN in J frame}
\begin{align}
G_\text{J}(\chi) &= \frac{G}{\varphi_0} \left( 1 + \frac{1}{2\omega_0 + 3} 
e^{- m_\text{J} \chi} \right)
\\
\gamma(\chi) &= \frac{1 - \frac{1}{2\omega_0 + 3} 
e^{-m_\text{J} \chi}}{1 + \frac{1}{2\omega_0 + 3} 
e^{-m_\text{J} \chi}}
\\
\begin{split}
\beta(\chi)
&= 1 + \frac{\varphi_0 \omega_1}{(2\omega_0+3)^3 
\left(1+\frac{e^{-m_\text{J} \chi}}{2\omega_0+3}\right)^2}
e^{-2 m_\text{J} \chi}
\\
&\qquad+ \frac{m_\text{J} \chi}{ 2 (2\omega_0+3)
\left(1+\frac{e^{-m_\text{J} \chi}}{2\omega_0+3}\right)^2}
\bigg[
2 e^{-m_\text{J} \chi} \ln(m_\text{J} \chi)
- e^{-2 m_\text{J} \chi}
- 2 \left( m_\text{J} \chi + e^{m_\text{J} \chi} \right) 
\text{Ei}(-2 m_\text{J} \chi)
\\
&\qquad+ \frac{3 \varphi_0}{2\omega_0+3} 
\left( \frac{U_3}{U_2} - \frac{1}{\varphi_0}
 - \frac{\omega_1}{2\omega_0 + 3} \right)
\left[ e^{-m_\text{J} \chi} \text{Ei}(-m_\text{J} \chi)
 - e^{m_\text{J} \chi} \text{Ei}(-3 m_\text{J} \chi) \right]
\bigg],
\end{split}
\end{align}
\end{subequations}
\end{widetext}
where 
\begin{align}
\label{equ:J frame mass}
m_\text{J} := \sqrt{\frac{2 U_2 \varphi_0}{2\omega_0+3}}
\end{align}
can be interpreted as the inverse range of the field or, roughly speaking, 
the mass of the field.
In this expression we use the notation $U_2 = U''(\varphi_0)/2$.
Here we make use of the exponential integral
\begin{align}
\text{Ei(-x)} := - \int_x^\infty da \frac{e^{-a}}{a}.
\end{align}


\subsection{PPN parameters in the Einstein frame}
\label{sec:PPN Parameter from the Einstein Frame}

In this section, we complement the Hohmann {\it et al.} \cite{hohmann.ppn} 
approach by calculating the PPN parameters for a general scalar-tensor theory
formulated in the Einstein-frame.
To do so, the equations of motion are solved order-by-order in 
the Einstein frame. Finally, we transform to the Jordan frame where the PPN 
parameters are defined. For the sake of understandability we perform the 
calculation in detail.

Here we consider a spacetime consisting of a point mass
surrounded by vacuum.
The stress-energy tensor is given by that
of a perfect fluid \cite{will.book}
\begin{align}
T^{\mu\nu} = \left( \rho + \rho \Pi + p \right) u^\mu u^\nu
+ p g^{\mu\nu},
\end{align}
with the rest-mass density $\rho$, the pressure $p$,
the specific energy density $\Pi$
and the four-velocity
$u^\mu$, satisfying $u_\mu u^\mu = -1$.
For solar system tests we typically have $\rho \gg p$ and $\rho \gg \rho \Pi$, 
so we may neglect both the effects of pressure and
specific energy density.
If the mass is at rest ($u^i = 0$), we obtain 
$T_{\mu\nu} = \text{diag}(\rho,0,0,0)$.
For a point source we have $\rho_\text{E} = M_\text{E} \delta(r) \epsilon$,
where the index $E$ implies that a quantity is defined in the Einstein frame.

For the metric in the Einstein frame we make the ansatz
\begin{align}
\begin{split}
ds_\text{E}^2 = &- \left( 1 - h_{\text{E}00}^{(1)}(r) \epsilon
 - h_{\text{E}00}^{(2)}(r) \epsilon^2 \right) dt_\text{E}^2 \epsilon^2
\\
&+ \left( 1 + h_{\text{E}rr}^{(1)}(r) \epsilon \right)
\left( dr^2 + r^2 d\Omega^2 \right),
\end{split}
\end{align}
where we choose isotropic coordinates with radial coordinate $r$.
We expand the scalar field in powers of $\epsilon$ and subsequently the 
coupling function and the potential are expanded around some constant 
value $\phi_0$:
\begin{subequations}
\label{equ expanding E frame quantities}
\begin{align}
\phi(r) &= \phi_0 + \phi_1(r) \epsilon + \phi_2(r) \epsilon^2
\\
F(\phi) &= F_0 + F_1 \left(\phi-\phi_0\right) + F_2 \left(\phi-\phi_0\right)^2
+ F_3 \left(\phi-\phi_0\right)^3
\\
V(\phi) &= V_0 + V_1 \left(\phi-\phi_0\right) + V_2 \left(\phi-\phi_0\right)^2
+ V_3 \left(\phi-\phi_0\right)^3
\end{align}
\end{subequations}

The left-hand sides of the equations \eqref{equ:STT EOMs E frame tensor},
the components of the Ricci tensor, are
\begin{subequations}
\begin{align}
\begin{split}
R_{00}^\text{E} &= -  \frac{1}{2} \nabla_r^2 h_{\text{E}00} \epsilon
- \frac{1}{2} \bigg( \nabla_r^2 h_{\text{E}00}^{(2)}
- h_{\text{E}rr}^{(1)}  \nabla_r^2 h_{\text{E}00}^{(1)}
\\&\quad+ \frac{1}{2}  ( \partial_r h_{\text{E}00}^{(1)})^2
+ \frac{1}{2} ( \partial_r h_{\text{E}00}^{(1)} ) 
( \partial_r h_{\text{E}rr}^{(1)} )\bigg) \epsilon^2
+ \mathcal{O}(\epsilon^3)
\end{split}
\\
R_{rr}^\text{E} &= \left( - \partial_r^2 h_{\text{E}rr}^{(1)} 
 - \frac{1}{r} \partial_r h_{\text{E}rr}^{(1)}
 + \frac{1}{2} \partial_r^2 h_{\text{E}00}^{(1)} \right) \epsilon
+ \mathcal{O}(\epsilon^2)
\\
R_{\theta\theta}^\text{E} &= \frac{1}{2} r^2 
\left( - \partial_r^2 h_{\text{E}rr}^{(1)}
- \frac{3}{r} \partial_r h_{\text{E}rr}^{(1)}
+ \frac{1}{r} \partial_r h_{\text{E}00}^{(1)} \right) \epsilon
+ \mathcal{O}(\epsilon^2)
\\
R_{\varphi\varphi}^\text{E} &= R_{\theta\theta}^\text{E} \sin^2\theta
+ \mathcal{O}(\epsilon^2)
\end{align}
\end{subequations}
to the required orders. All other components are identically zero.
The right-hand sides are
\begin{align}
\begin{split}
R_{\mu\nu}^\text{E}
&= \frac{1}{2} \eta_{\mu\nu} V_0
+ \bigg[ 8 \pi G \left( \delta_\mu^0 \delta_\nu^0
 + \frac{1}{2} \eta_{\mu\nu} \right) M_\text{E} \delta(r)
\\
&\quad+ \frac{1}{2} \left( h_{\text{E}\mu\nu}^{(1)} V_0
 + \eta_{\mu\nu} V_1 \phi_1 \right) \bigg] \epsilon
\\
&\quad+ \bigg[ \frac{8 \pi G}{2} \left( h_{\text{E}\mu\nu}^{(1)}
 + \eta_{\mu\nu} h_{\text{E}00}^{(1)} \right) M_\text{E} \delta(r)
\\
&\quad+ 2 \partial_r \phi_1 \partial_r \phi_1 \delta_\mu^r \delta_\nu^r
+ \frac{1}{2} \eta_{\mu\nu} V_2 \phi_1^2
\\
&\quad+ \frac{1}{2} (h_{\text{E}\mu\nu}^{(1)} \phi_1
 + \eta_{\mu\nu} \phi_2 ) V_1
\bigg] \epsilon^2 + \mathcal{O}(\epsilon^3).
\end{split}
\end{align}
The flat-space Minkowski metric is, with our choice of coordinates, 
$\eta_{\mu\nu} = \text{diag}(-1,1,r^2,r^2 \sin^2\theta)$.
Calculating both sides of the scalar equation yields
\begin{align}
\begin{split}
\nabla_\text{E}^2 \phi
&= \nabla_r^2 \phi_1 \epsilon
+ \bigg[ \nabla_r^2 \phi_2 - h_{\text{E}rr}^{(1)} \nabla_r^2 \phi_1
\\&\quad+ \frac{1}{2} \left( \partial_r h_{\text{E}rr}^{(1)}
 - \partial_r h_{\text{E}00}^{(1)} \right) \partial_r \phi_1
\bigg] \epsilon^2
\end{split}
\end{align}
and
\begin{align}
\begin{split}
\nabla_\text{E}^2 \phi &=
\frac{1}{4} V_1
+ \left[-\frac{F_1}{F_0} 2 \pi G M_\text{E} \delta(r)
 + \frac{1}{2} V_2 \phi_1 \right] \epsilon 
\\
&\quad + \bigg[ \bigg( 2 \pi G M_\text{E} 
\left( \frac{F_1^2}{F_0^2}-\frac{2 F_2}{F_0} \right) \phi_1
\\
&\quad - 2 \pi G M_\text{E} \frac{F_1}{F_0} 
h_{\text{E}00}^{(1)} \bigg) \delta(r)
+ \frac{3}{4} V_3 \phi_1^2 + \frac{1}{2} V_2 \phi_2
\bigg] \epsilon^2.
\end{split}
\end{align}
By $\nabla_r^2$ we mean the flat space spherical coordinate Laplace operator,
$\nabla_r^2 := \partial_r^2 + 2/r \partial_r$.
First, we consider the zeroth-order equations
\begin{subequations}
\begin{align}
&0 = \frac{1}{2} \eta_{\mu\nu} V_0
\\
&0 = \frac{1}{4} V_1,
\end{align}
\end{subequations}
which require $V_0=V_1=0$. At first order in $\epsilon$, the scalar equation is
\begin{align}
\label{equ:scalar eom}
\left(\nabla_r^2 - m_\text{E}^2 \right) \phi_1 =
 -\frac{F_1}{F_0} 2 \pi G M_\text{E} \delta(r),
\end{align}
with solution
\begin{align}
\phi_1(r) = \frac{F_1}{2 F_0} \frac{G M_\text{E}}{r} e^{-m_\text{E} r},
\end{align}
where we have defined
\begin{align}
m_\text{E}^2 := \frac{1}{2} V_2.
\end{align}
The first order $00$-tensor equation and its solution are
\begin{align}
\begin{split}
-  \frac{1}{2} \nabla_r^2 h_{\text{E}00}
&= 8 \pi G \frac{1}{2} M_\text{E} \delta(r)
\\
\rightarrow \quad h_{\text{E}00}(r) &= \frac{2 G M_\text{E}}{r} .
\end{split}
\end{align}
At the same order, the $rr$-equation is
\begin{align}
- \partial_r^2 h_{\text{E}rr}^{(1)} 
 - \frac{1}{r} \partial_r h_{\text{E}rr}^{(1)}
 + \frac{1}{2} \partial_r^2 h_{\text{E}00}^{(1)}
= 8 \pi G \frac{1}{2} M_\text{E} \delta(r)
\end{align}
and the $\theta\theta$-equation turns into
\begin{align}
- \partial_r^2 h_{\text{E}rr}^{(1)}
- \frac{3}{r} \partial_r h_{\text{E}rr}^{(1)}
+ \frac{1}{r} \partial_r h_{\text{E}00}^{(1)}
= 8 \pi G  M_\text{E} \delta(r).
\end{align}
Summing these two equations yields
\begin{align}
\nabla_r^2 h_{\text{E}rr}^{(1)} = -8 \pi G  M_\text{E} \delta(r),
\end{align}
with solution
\begin{align}
h_{\text{E}rr}^{(1)}(r) = \frac{2 G M_\text{E}}{r}.
\end{align}
The $00$-tensor equation at second order turns into
\begin{align}
\begin{split}
\nabla_r^2 h_{\text{E}00}^{(2)}
=
&- \frac{4 G^2 M_\text{E}^2}{r^4}
+ \frac{F_1^2}{4 F_0^2} V_2 \frac{G^2 M_\text{E}^2}{r^2} e^{-2 m_\text{E} r^2},
\end{split}
\end{align}
where we dropped a term proportional to $\delta(r) h_{\text{E}rr}^{(1)}$ since
it corresponds to gravitational self-energy \cite{hohmann.ppn} and we get the
solution
\begin{align}
\begin{split}
h_{\text{E}00}^{(2)}(r)
= - \frac{4 G^2 M_\text{E}^2}{2 r^2}
\bigg[ &1 - \frac{F_1^2}{4 F_0^2}
\bigg( \frac{1}{2} m_\text{E} r e^{-2 m_\text{E} r}
\\&+ m_\text{E}^2 r^2 \text{Ei}(- 2 m_\text{E} r) \bigg)  \bigg].
\end{split}
\end{align}
We notice that the metric component at post-Newtonian order has an additional
term compared to the Schwarzschild metric of general relativity.
The second order scalar field equation is
\begin{align}
\begin{split}
\nabla_r^2 \phi_2 - h_{\text{E}rr}^{(1)} \nabla_r^2 \phi_1
&+ \frac{1}{2} \left( \partial_r h_{\text{E}rr}^{(1)}
 - \partial_r h_{\text{E}00}^{(1)} \right) \partial_r \phi_1
\\&= \frac{3}{4} V_3 \phi_1^2 + \frac{1}{2} V_2 \phi_2.
\end{split}
\end{align}
Also here, we dropped the gravitational self-energy terms proportional to 
$\phi_1 \delta(r)$, $h_{\text{E}00}^{(1)} \delta(r)$ and 
$h_{\text{E}rr}^{(1)} \delta(r)$.
As solution we find
\begin{align}
\begin{split}
\phi_2(r) &= \frac{1}{4} \frac{F_1}{2 F_0} m_\text{E} 
 \frac{4 G^2 M_\text{E}^2}{r}
\\&\qquad\times \left[ e^{m_\text{E} r} \text{Ei}(-2 m_\text{E} r)
 - e^{-m_\text{E} r} \ln(m_\text{E} r) \right]
\\
&\quad+ \frac{1}{2 m_\text{E}} \frac{3 F_1^2}{64 F_0^2} V_3
 \frac{4 G^2 M_\text{E}^2}{r}
\\&\qquad\times \left[ e^{m_\text{E} r} \text{Ei}(-3 m_\text{E} r)
 - e^{-m_\text{E} r} \text{Ei}(-m_\text{E} r)  \right].
\end{split}
\end{align}

We have thus solved the equations of motion to post-Newtonian order. 
To determine the PPN parameters we must turn to the Jordan frame where 
they are defined. The metric line elements in the two frames are related 
by the conformal transformation \eqref{equ:conformal transformation}, giving
\begin{align}
\begin{split}
ds^2_\text{J} &= F(\phi)^{-1} ds^2_\text{E}
\\
&= -\bigg[ 1 - \left(h_{\text{E}00}^{(1)} + \frac{F_1}{F_0} \phi_1 \right)
 \epsilon
- \bigg( h_{\text{E}00}^{(2)} - \frac{F_1}{F_0} h_{\text{E}00}^{(1)} \phi_1
\\&\quad+ \left( \frac{F_2}{F_0} - \frac{F_1^2}{F_0^2} \right) \phi_1^2
 + \frac{F_1}{F_0} \phi_2 \bigg) \epsilon^2 \bigg] \frac{dt_\text{E}^2}{F_0},
\\
&\quad +\left[ 1 + \left( h_{\text{E}rr}^{(1)} - \frac{F_1}{F_0} \phi_1 \right)
 \epsilon \right] \left( \frac{dr^2}{F_0} + \frac{r^2}{F_0} d\Omega^2 \right).
\end{split}
\end{align}
Comparing this to the metric in the Jordan frame 
\eqref{equ Jordan frame metric ansatz}, we find
\begin{subequations}
\begin{align}
h_{\text{J}00}^{(1)} &= \frac{2 G_\text{J} M_\text{J}}{\chi}
\stackrel{!}{=} h_{\text{E}00}^{(1)} + \frac{F_1}{F_0} \phi_1
\\
h_{\text{J}\chi\chi}^{(1)} &= \gamma(\chi)\frac{2 G_\text{J} M_\text{J}}{\chi}
\stackrel{!}{=} h_{\text{E}rr}^{(1)} - \frac{F_1}{F_0} \phi_1
\\
\begin{split}
h_{\text{J}00}^{(2)} &=
 - \beta(\chi) \frac{4 G_\text{J}^2 M^2_\text{J}}{2\chi^2}
\\
&\stackrel{!}{=} h_{\text{E}00}^{(2)}
 - \frac{F_1}{F_0} h_{\text{E}00}^{(1)} \phi_1
 + \left( \frac{F_2}{F_0} - \frac{F_1^2}{F_0^2} \right) \phi_1^2
 + \frac{F_1}{F_0} \phi_2
\end{split}
\end{align}
\end{subequations}
with
\begin{subequations}
\begin{align}
t_\text{J} &= \frac{t_\text{E}}{\sqrt{F_0}}
\\
\chi &= \frac{r}{\sqrt{F_0}}.
\end{align}
\end{subequations}
From the $h_{\text{J}00}^{(1)}$ relation we can identify the effective 
gravitational `constant' in the Jordan frame
\begin{align}
\begin{split}
G_\text{J}(r)
&= \frac{r}{2 F_0 M_\text{E}} \left( h_{\text{E}00}^{(1)}
 + \frac{F_1}{F_0} \phi_1 \right),
\end{split}
\end{align}
where the masses in the Jordan frame satisfy
\begin{subequations}
\begin{align}
m_\text{J} &= \sqrt{F_0} m_\text{E}
\\
M_\text{J} &= \sqrt{F_0} M_\text{E},
\end{align}
\end{subequations}
such that $m_\text{J} \chi = m_\text{E} r$.
With this we obtain the $\gamma$ parameter
\begin{align}
\begin{split}
\gamma(r)
&= \frac{h_{\text{E}rr}^{(1)} - \frac{F_1}{F_0} \phi_1}{ h_{\text{E}00}^{(1)}
 + \frac{F_1}{F_0} \phi_1 }.
\end{split}
\end{align}
And finally, the $\beta$ parameter is
\begin{widetext}
\begin{align}
\begin{split}
\beta(r)
&= \frac{2\chi^2}{4 G_\text{J}^2(r) M^2_\text{J}}
\left[ \frac{F_1}{F_0} h_{\text{E}00}^{(1)} \phi_1
- \left( \frac{F_2}{F_0} - \frac{F_1^2}{F_0^2} \right) \phi_1^2
- \frac{F_1}{F_0} \phi_2
- h_{\text{E}00}^{(2)}
 \right]
\\
&=1 + \frac{1}{\left( h_{\text{E}00}^{(1)} + \frac{F_1}{F_0} \phi_1 \right)^2}
\left[ \left( \frac{F_1^2}{F_0^2} 
- \frac{2 F_2}{F_0} \right) \phi_1^2
- \left( h_{\text{E}00}^{(1)} \right)^2
- \frac{2 F_1}{F_0} \phi_2
- 2 h_{\text{E}00}^{(2)} 
\right].
\end{split}
\end{align}
\end{widetext}

Inserting the scalar field and metric components determined above, we obtain 
the PPN parameters for a scalar-tensor theory formulated in the Einstein frame:
\begin{widetext}
\begin{subequations}
\begin{align}
G_\text{J}(r) &= \frac{G }{ F_0} 
\left( 1 + \frac{F_1^2}{4 F_0^2} e^{-m_\text{E} r} \right)
\\
\gamma(r) &=\frac{ 1 - \frac{F_1^2}{4 F_0^2} e^{-m_\text{E} r}  }
{ 1 + \frac{F_1^2}{4 F_0^2} e^{-m_\text{E} r} }.
\\
\begin{split}
\beta(r) &= 1 + \frac{F_1^2}{4 F_0^2 \left( 1 + \frac{F_1^2}{4 F_0^2} 
e^{-m_\text{E} r} \right)^2}
\left( \frac{F_1^2}{4 F_0^2} - \frac{F_2}{2 F_0} \right) e^{- 2 m_\text{E} r }
\\
&\qquad+ \frac{F_1^2 m_\text{E} r}{32 F_0^2 
\left( 1 + \frac{F_1^2}{4 F_0^2} e^{-m_\text{E} r} \right)^2}
\bigg[
8 e^{-m_\text{E} r} \ln{m_\text{E} r}
- 4 e^{- 2 m_\text{E} r }
- 8 \left( e^{m_\text{E} r} + m_\text{E} r \right) \text{Ei}(-2m_\text{E} r)
\\
&\qquad\qquad\qquad\qquad\qquad\qquad\qquad
+ 3 \frac{F_1}{F_0} \frac{V_3}{V_2} 
\left( e^{-m_\text{E} r} \text{Ei}(-m_\text{E} r) - e^{m_\text{E} r} 
\text{Ei}(-3m_\text{E} r) \right)
\bigg]
\end{split}
\end{align}
\end{subequations}
\end{widetext}

To compare this result to the one found in \cite{hohmann.ppn},
we use the transformation laws for the coupling functions 
\eqref{equ coupling functions relations E J frame} and the potentials 
\eqref{equ potentials relations E J frame}. Indeed, this leads to equations 
\eqref{equ:PPN in J frame}.

If we choose to neglect the second order deviation from the Schwarzschild 
metric (i.e. $h_{\text{E}00}^{(2)} = - 4 G^2 M_\text{E}^2/(2r^2)$) and only 
consider the leading order scalar field contribution (i.e. we set $\phi_2=0$), 
then the effective coupling constant and the PPN parameters simplify to
\begin{subequations}
\label{equ ppn bundle for general field in E frame}
\begin{align}
G_\text{J}(r) &= \frac{G}{F_0 } \left( 1 + \frac{F_1}{2 F_0} 
\frac{r}{G M_\text{E}} \phi_1 \right)
\\
\gamma(r) &= \frac{ 1 - \frac{F_1}{2 F_0} \frac{r}{G M_\text{E}} \phi_1}
{  1 + \frac{F_1}{2 F_0} \frac{r}{G M_\text{E}} \phi_1 }
\\
\beta(r)
&= 1 + \frac{\left( \frac{F_1^2}{4 F_0^2} - \frac{F_2}{2 F_0} \right) 
\phi_1^2}{\left( \frac{G M_\text{E}}{r} + \frac{F_1}{2 F_0} \phi_1 \right)^2}.
\end{align}
\end{subequations}
We notice that on the one hand, any non-trivial scalar-tensor theory predicts 
a $\gamma$ different from its general relativity value $1$. On the other hand, 
it is still possible to have $\beta = 1$: if $F_1^2/F_0 - 2 F_2 = 0$. This 
condition is equivalent to $F'(\phi_0)^2/F(\phi_0) - F''(\phi_0) = 0$ which 
is solved by $F(\phi) = c_1 \exp(c_2 \phi)$. An exponential coupling function 
in the Einstein frame corresponds to a constant coupling function in the Jordan
frame, $\omega = \omega_0$, and therefore to a Brans-Dicke-like theory as for 
instance the original chameleon model (see sections \ref{sec:Brans-Dicke} and 
\ref{subsec:Chameleon Theory}).


In the following section we discuss current constraints on the PPN parameters 
and apply them to our formalism. In particular the constraint on $\gamma$ 
coming from the Cassini spacecraft is discussed. This is followed by some 
important examples of scalar-tensor theories.

\section{Experimental framework and constraints}
\label{Framework to Constrain PPN Parameters}

Above we have discussed the scalar field in vacuum outside a point source in 
the weak field limit. The assumption of a point source is obviously not correct
for experiments performed around extended objects as the Earth or the Sun.
Within such an object the field can behave very different to that of a point 
source and screening mechanisms can show up due to non-linear effects.
But still, since the density in the solar system is very low, one can 
expect the field to maintain the form $\phi_1 \sim e^{-m r}/r$ in the low 
density region outside some source. 
Therefore, we make the ansatz
\begin{subequations}
\label{equ general scalar field profile}
\begin{align}
& \varphi(\chi) = \varphi_0 + \varphi_1(\chi) = \varphi_0 
+ \xi \frac{2}{2\omega_0+3} \frac{G M_\text{J}}{\chi} 
e^{-m_\text{J} (\chi-X)}
\\
& \phi(r) = \phi_0 + \phi_1(r) = \phi_0 + \xi \frac{F_1}{2 F_0} 
\frac{G M_\text{E}}{r} e^{-m_\text{E} (r-R)}
\end{align}
\end{subequations}
for the exterior field up to first order, written in the Jordan and the 
Einstein frame, respectively. 
By $X$ (Jordan frame) and $R=\sqrt{F_0}X$ (Einstein frame) we 
denote the size of the object.
Therefore, the field starts falling off exponentially at the surface of the
source instead of at its center.
Notice that we introduced some arbitrary parameter $\xi$.
By doing so, we are able to discuss constraints on the PPN parameters around
more realistic sources, also for theories containing screening mechanisms
without knowing their exact nature.
The $\xi$ parameter describes how much the exterior field deviates from 
that generated by a point source ($\xi = 1$) with the same mass.
In other words, a source can act as an effective point source 
of mass $\xi M$.
In particular, we will show in section \ref{subsec:Massive Brans-Dicke}
that a massive Brans-Dicke scalar field
takes this form if we consider the source to be a sphere
with constant density. We find an expression for $\xi$  which
depends on both the mass of the scalar field and the radius 
of the source.

Plugging the ansatz above into 
\eqref{equ ppn bundle for general field in E frame} 
we find the effective gravitational constant and the PPN parameters
\begin{subequations}
\label{equ ppn bundle for constraint bundle}
\begin{align}
\begin{split}
G_\text{J} &= \frac{G}{F_0 } \left( 1 + \xi \frac{F_1^2}{4 F_0^2} 
e^{-m_\text{E} (r-R)} \right)
\\&= \frac{G}{\varphi_0 } \left( 1 + \xi \frac{1}{2\omega_0+3} 
e^{-m_\text{J} (\chi-X)} \right)
\end{split}
\\
\gamma &= \frac{ 1 - \xi \frac{F_1^2}{4 F_0^2} e^{-m_\text{E} (r-R)} }
{1 + \xi \frac{F_1^2}{4 F_0^2} e^{-m_\text{E} (r-R)} }
= \frac{ 1 - \frac{\xi}{2\omega_0+3} e^{-m_\text{J} (\chi-X)} }
{1 + \frac{\xi}{2\omega_0+3} e^{-m_\text{J} (\chi-X)} }
\label{equ:gamma parameter for constraint}
\\
\begin{split}
\beta
&= 1 + \frac{\left(1 - \frac{2 F_0 F_2}{F_1^2} \right) }
{ \left[ 1 + \left( \xi \frac{F_1^2}{4 F_0^2} e^{-m_\text{E} (r-R)}
\right)^{-1} \right]^2}
\\
&= 1 + \frac{\frac{\varphi_0 \omega_1}{2\omega_0 + 3}}
{ \left[ 1 + \left( \xi \frac{1}{2\omega_0 + 3} e^{-m_\text{J} (\chi-X)}
\right)^{-1} \right]^2}.
\end{split}
\end{align}
\end{subequations}

Typically, experimental constraints on PPN parameters are used to limit the
$(\omega_0,\tilde{m}_\text{J})$-parameter space \cite{hohmann.ppn,periv}.
The definition of the $\omega_0$-independent mass 
\begin{align}
\tilde{m}_\text{J} := \sqrt{2 U_2 \varphi_0}
\end{align}
is required in order to have two independent parameters since the 
original mass $m_\text{J}$, defined in \eqref{equ:J frame mass},
depends on $\omega_0$.
But since we want to incorporate possible screening mechanisms in 
extended sources, giving us the additional parameter $\xi$,
we consider a slightly different approach. We define a new parameter
\begin{align}
\alpha := \frac{\xi}{2\omega_0+3} = \xi \frac{F_1^2}{4 F_0^2},
\end{align}
allowing us to constrain the $(\alpha,m_\text{J})$-parameter space.
Notice that $\alpha$ contains two different kinds of parameters. 
First, the components of the scalar coupling functions, 
$\omega_0$ and $F_1^2/F_0^2$, depend on the underlying theory 
of gravity only and are the same everywhere.
Second, the parameter $\xi$ can depend on properties 
of the source, as its composition.
Therefore, it can vary drastically among different sources.

\begin{figure*}
\includegraphics{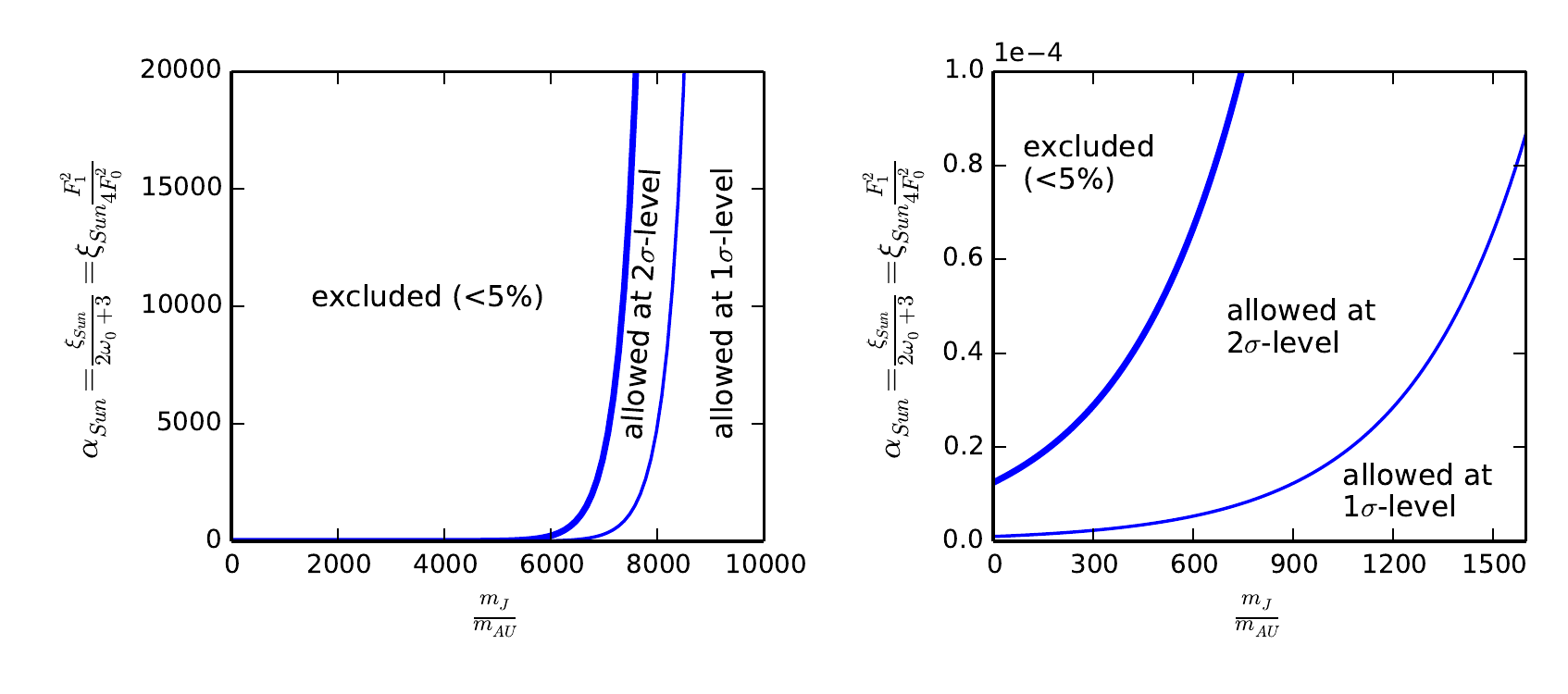}
\caption{Cassini constraint on scalar-tensor theories.
The Cassini constraint $\gamma = 1+(2.1\pm 2.3)\cdot 10^{-5}$ 
(at $1\sigma$-level) together with 
equation \eqref{equ:gamma parameter for constraint} and 
$\chi-X_\text{Sun} = 0.6 \text{ solar radii} = 0.00279 \text{AU}$ are used to
constrain the $(\alpha_\text{Sun},m_\text{J})$-parameter space.
The solid lines divide the plots into regions that are excluded
(probability $< 5\%$) and that are allowed at the $2\sigma$-level
and at the $1\sigma-$level, respectively.
The x-axis shows the mass, i.e. the inverse range, 
of the scalar field in the Jordan 
frame in terms of inverse astronomical units $m_\text{AU} = 1/\text{AU}$, 
the y-axis shows $\alpha_\text{Sun} = \xi_\text{Sun}/(2\omega_0+3)
= \xi_\text{Sun} F_1^2/(4 F_0^2)$ where 
$\xi_\text{Sun}$ is a parameter characteristic for the Sun.
}
\label{fig:Cassini constraint}
\end{figure*}

There are different experimental constraints on the PPN parameters.
The most stringent one comes from measuring the 
frequency shift of a radio signal sent from and to the Cassini spacecraft 
while close to conjunction with the Sun, with 
$\gamma = 1+(2.1\pm 2.3)\cdot 10^{-5}$ at the $1\sigma$-confidence level 
\cite{cassini.nature}. 
The closest distance between the propagating signal and the center
of the Sun was $1.6$ solar radii.
We can now use this to constrain the parameter space 
$(\alpha_\text{Sun},m_\text{E})$,
as shown in figure \ref{fig:Cassini constraint}.

The perihelion precession of Mercury gives the constraint 
$|2\gamma - \beta -1| < 3\cdot 10^{-3}$ \cite{will.review}.
Planetary ephemerides are used to constrain $|\gamma-1|$ and $|\beta-1|$ to the
$10^{-5}$-level \cite{fienga.2014,pitjeva}. But since the gravitational 
interaction does not take place at a fixed distance from some massive body, 
this limit cannot be used to constrain the distance dependent parameters 
discussed here. 

Scalar theories can also be constrained by accurate measurements of the periods
of binary pulsars: if scalar radiation is emitted, it results in a change of 
the orbital period \cite{pulsar.brax}.

The GAIA mission launched in $2013$, located at the Sun-Earth Lagrange point 
L2, is expected to improve the constraint on $\gamma$ to the $10^{-6}$ level 
\cite{gaia.gamma} via relativistic astrometry by precisely monitoring the 3D 
motion of planets and stars in our galaxy.

In the following subsections, we consider specific theories of gravity and 
use the above formalism to calculate their PPN parameters. 
The Cassini measurement can then be used to constrain these theories. 
As atomic clocks become more accurate, clock carrying satellites that orbit 
the Earth will place constraints on the value of the 
PPN parameters around our own planet. We will discuss such measurements in 
Sec.\ \ref{Constraining PPN Parameters in Earth Orbit}.


\subsection{Brans-Dicke theory}
\label{sec:Brans-Dicke}

The simplest example and the prototype of scalar-tensor theories is Brans-Dicke
theory \cite{brans.dicke}. In the Jordan frame it is defined to have a constant
coupling $\omega = \omega_0$ and a vanishing scalar potential, leaving
the field massless, $m_\text{J} = 0$. Therefore, the PPN parameters will not 
have a distance dependence and we have $\xi = 1$ because no hiding mechanism 
occurs.

In this theory, $\omega_0$ is the only parameter. With 
\eqref{equ:relation E and J frame fields} and 
\eqref{equ:E frame coupling expanded} we obtain the coupling function 
$F(\phi) = F_0 \exp[\pm 2(\phi-\phi_0)/\sqrt{2\omega_0+3}]$.
Therefore, in the Einstein frame, Brans-Dicke theories have an exponential 
coupling function with no scalar potential term. This gives $\beta = 1$ as 
in general relativity and $\gamma = (\omega_0+1)/(\omega_0+2)$. Using the 
Cassini constraint on $\gamma$, one finds that $\omega_0>40'000$ at the 
$2\sigma$-level.


\subsection{Eddington-Robertson metric}
\label{sub:Eddington Robertson Metric}

Assuming that the potential vanishes $U=V=0$ and then solving the equations 
of motion yields the PPN parameters
\begin{subequations}
\begin{align}
G_\text{J} &= \frac{G}{\varphi_0} \left( 1 + \frac{1}{2\omega_0 + 3} \right)
 = \frac{G }{ F_0} \left( 1 + \frac{F_1^2}{4 F_0^2} \right)
\\
\gamma &= \frac{1 + \omega_0}{2+\omega_0} = 
\frac{ 1 - \frac{F_1^2}{4 F_0^2} }{ 1 + \frac{F_1^2}{4 F_0^2} }
\\
\beta &= 1 + \frac{\varphi_0 \omega_1 }{ (2\omega_0 + 3) (2\omega_0 + 4)^2} 
= 1 + \frac{ \frac{F_1^2}{4 F_0^2} \left( \frac{F_1^2}{4 F_0^2}
 - \frac{F_2}{2 F_0} \right) }{\left( 1 + \frac{F_1^2}{4 F_0^2} \right)^2}.
\end{align}
\end{subequations}
Due to the absence of the potential there is no distance dependence in both
$G_\text{J}$ and the PPN parameters.
The metric \eqref{equ Jordan frame metric ansatz} with these constant 
parameters was given by Eddington and Robertson \cite{weinberg.book}. 
So, for fixed values of $\gamma$ and $\beta$ we can invert this expressions
to obtain the components of the coupling function in the Jordan frame
\begin{subequations}
\begin{align}
\omega_0 &= -\frac{2\gamma - 1}{\gamma-1}
\\
\omega_1 &= - \frac{4 (\beta-1) (\gamma+1) }{\varphi_0 (\gamma-1)^3}
\end{align}
\end{subequations}
and accordingly in the Einstein frame
\begin{subequations}
\begin{align}
\frac{F_1^2}{4 F_0^2} &= \frac{1-\gamma}{1+\gamma}
\\
\frac{F_2}{2 F_0} &= \frac{5 - 4\beta - 2\gamma + \gamma^2}{1-\gamma^2}.
\end{align}
\end{subequations}


\subsection{Massive Brans-Dicke Theory}
\label{subsec:Massive Brans-Dicke}

Here we solve the scalar field equation of a massive
Brans-Dicke scalar field generated by a more
realistic source than a point mass.
We consider a coupling function which is, 
as in the original Brans-Dicke theory,
exponential in the Einstein frame and thus
constant ($\omega = \omega_0$) in the Jordan frame.
Further, in the Einstein frame we add a quadratic potential 
$V = V_2 (\phi - \phi_0)^2$ with $V_2 = 2 m_\text{E}$.
This corresponds to the potential $U = U_2(\varphi - \varphi_0)^2$
in the Jordan frame.

For the case of a point source, constraints on massive
Brans-Dicke fields have been discussed
in \cite{periv, hohmann.ppn}.
There, the authors used the Cassini constraint on $\gamma$ to
limit the $(\tilde{m}_\text{J},\omega_0)$-parameter space.
Here, we extend this discussion by replacing the point source with
a more realistic density distribution.
This will allow us to determine the parameter $\xi$, introduced in
\eqref{equ general scalar field profile}.

We consider a static spherically symmetric mass with radius $R$ and constant
density $\rho_{E0}$ (i.e. $\rho_{E}(r)=\rho_{E0}$ for $r<R$ and 
$\rho_{E}(r)=0$ otherwise)
and we neglect the gravitational effects of pressure.
Further we assume that the mass is surrounded by vacuum.
The equation of motion is given by
\begin{align}
\left(\nabla_r^2 - m_\text{E}^2 \right) \phi_1(r) =
 -\frac{F_1}{F_0} 2 \pi G \rho_\text{E}(r),
\end{align}
which follows from \eqref{equ:scalar eom}.
To solve this equation we make use of the Green function
$G(\vec{r}) = -e^{-m_\text{E}r}/4\pi r$,
solving the equation
$\left(\nabla_r^2 - m_\text{E}^2 \right) G(\vec{r}) = \delta(\vec{r})$.
Then we find the scalar field by integrating
\begin{align}
\begin{split}
\phi_1(\vec{r}) &= \int G(|(\vec{r})-(\vec{s})|) \left( -\frac{F_1}{F_0}
2 \pi G \rho_\text{E}(s) \right) d^3\vec{s}
\\
&= \frac{F_1}{F_0} \pi G \rho_{E0} 
\\
&\quad \times \int_0^\pi \int_0^R 
\frac{e^{-m_\text{E} \sqrt{r^2+s^2-2rs\cos\theta }}}
{\sqrt{r^2+s^2-2rs\cos\theta }} s^2 \sin\theta \, ds \, d\theta.
\end{split}
\end{align}
To obtain the exterior solution $\phi_1^\text{ext}(r>R)$, the integrand is
expanded around $s/r = 0$, since $r>s$ for all $s$.
This allows each term of the Taylor series to be integrated, giving
\begin{align}
\begin{split}
\phi_1^\text{ext}(r>R) = &\frac{F_1}{F_0} \frac{\pi G \rho_{E0}}{r} 
e^{-m_\text{E} r}
\\
&\times \sum_{k=0}^{\infty} m_\text{E}^{2k} R^{2k+3} \frac{2}{(2k+1)! (2k+3)}.
\end{split}
\end{align}
Finally, this can be written as
\begin{align}
\label{equ: massive BD exterior solution}
\begin{split}
\phi_1^\text{ext}(r>R) = &\left(3 \frac{m_\text{E} R \cosh (m_\text{E} R)
 - \sinh(m_\text{E} R)}{ R^3 m_\text{E}^3}
e^{-m_\text{E} R} \right)
\\
&\times \frac{F_1}{2 F_0} \frac{G M_\text{E}}{r} e^{-m_\text{E}(r-R)},
\end{split}
\end{align}
where in the last step we substituted 
$M_\text{E} = 4\pi/3 \rho_{E0} R^3$.

The interior solution $\phi_1^\text{ext}(r<R)$ is obtained by splitting the
integral over $s$ into two parts. 
First, we perform the integral $\int_0^r$ where we can expand around
$s/r=0$ and find the solution analogous to the exterior solution.
Second, for the integral from $\int_r^R$ we can expand around
$r/s=0$. Together, we find
\begin{align}
\begin{split}
\phi_1^\text{int}(r<R) = & 3 \frac{F_1}{2 F_0} \frac{G M_\text{E}}{R^3}
\bigg[\frac{e^{-m_\text{E} r}}{m_\text{E}^2}
\left( \cosh{(m_\text{E} r)} + \sinh{(m_\text{E} r)} \right)
\\
&- e^{-m_\text{E} R} \frac{1+m_\text{E} R}{m_\text{E}^3 r} 
\sinh{(m_\text{E} r)} \bigg].
\end{split}
\end{align}

Notice that the exterior solution \eqref{equ: massive BD exterior solution}
precisely coincides with the general ansatz
\eqref{equ general scalar field profile} if we choose
\begin{align}
\label{equ: xi for massive BD}
\xi = 3 \frac{m_\text{E} R \cosh (m_\text{E} R)
 - \sinh(m_\text{E} R)}{ R^3 m_\text{E}^3}
e^{-m_\text{E} R}.
\end{align}

The solution expressed in the Jordan frame is
\begin{align}
\begin{split}
\varphi_1^\text{ext}(\chi>X) = &\left(3 \frac{m_\text{J}
 X \cosh (m_\text{J} X) - \sinh(m_\text{J} X)}{ X^3 m_\text{J}^3}
e^{-m_\text{J} X} \right)
\\
&\times \frac{2}{2 \omega_0+3} \frac{G M_\text{J}}{\chi}
 e^{-m_\text{J}(\chi-X)}
\\
\varphi_1^\text{int}(\chi<X) = & \frac{6}{2 \omega_0+3}
 \frac{G M_\text{J}}{X^3}
 \\
& \times \bigg[\frac{e^{-m_\text{J} \chi}}{m_\text{J}^2} 
\left( \cosh{(m_\text{J} \chi)} + \sinh{(m_\text{J} \chi)} \right)
\\
&\qquad- e^{-m_\text{J} X} \frac{1+m_\text{J} X}{m_\text{J}^3 \chi} 
\sinh{(m_\text{J} \chi)} \bigg].
\end{split}
\end{align}

Since the $\gamma$ parameter
\eqref{equ:gamma parameter for constraint}
depends on $\xi$, it depends on the size of the source.
In the massless limit $m_\text{E/J} \rightarrow 0$, 
$\xi$ approaches unity.
Then, $\gamma$ depends on properties of the theory only
and is independent of $R$.
In the limit of vanishing radius, $\xi$ approaches unity as well, 
giving the same result as for a point source.

While in \cite{periv} the interaction distance is assumed to be 
$r = 1\,\text{AU}$, 
Hohmann {\it et al.} \cite{hohmann.ppn} choose $r = 1.6$ solar radii 
since this is the closest
distance between the signal and the Sun.
This dramatically improves the constraints on 
$\tilde{m}_\text{J}$ and $\omega_0$.
Including $\xi$ given by \eqref{equ: xi for massive BD}, which accounts
for the assumption that the Sun is a sphere with constant density, 
the constraint on the $(\tilde{m}_\text{J},\omega_0)$-parameter space given
by the Cassini experiment is shown in figure 
\ref{fig:massive BD} (solid lines),
where we assume that $r = 1.6$ solar radii. 
\begin{figure*}
\centering
\includegraphics{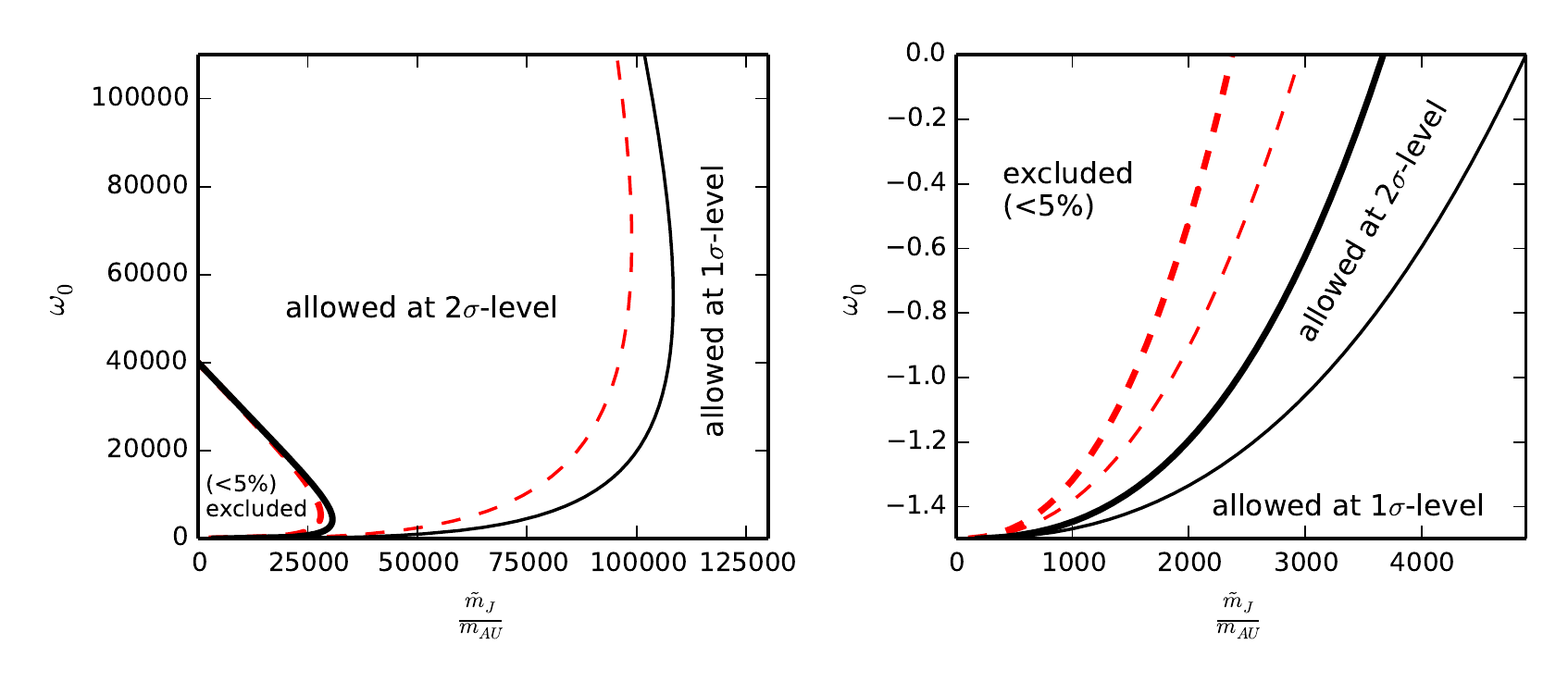}
\caption{\label{fig:massive BD} 
Cassini constraint on massive Brans-Dicke theory.
The constraint on PPN $\gamma$ given by the Cassini experiment is
used to constrain the $(\tilde{m}_\text{J},\omega_0)$-parameter space
for massive Brans-Dicke theory.
For this theory, the parameter $\xi$ in the expression for
$\gamma$ is given by \eqref{equ: xi for massive BD}.
For the interaction distance we take 
$\chi-X_\text{Sun} = 0.6 \text{ solar radii}$.
The solid lines separate the regions which are excluded 
(probability $< 5\%$), that are allowed at the
$2\sigma-$level and at the $1\sigma-$level, respectively.
The dashed-lines show the corresponding boundaries
between these regions for the case where the Sun
is considered to be a point mass.
The x-axis shows the $\omega_0$-independent mass 
$\tilde{m}_\text{J}$ in terms of inverse astronomical units 
$m_\text{AU} = 1/\text{AU}$, the y-axis shows $\omega_0$.
}
\label{fig:massive BD}
\end{figure*}
Comparing to the dashed lines which represent the analogous
result for a point source, we notice that the constraints are
more stringent if an extended source is considered.
This is due to the fact that, even though $\xi < 1$, the field falls
off like $e^{-m_\text{E}(r-R)}$ instead of $e^{-m_\text{E}r}$.


\subsection{Chameleon theory}
\label{subsec:Chameleon Theory}

Another example of a class of scalar-tensor theories are chameleon theories, 
introduced by Khoury and Weltman 
\cite{chameleon.khoury.prl,*chameleon.khoury}. 
They allow a very
light cosmological scalar field that couples to matter with gravitational 
strength and satisfies current observational constraints. Formulated in the 
Einstein frame, chameleons have, as Brans-Dicke theory does, an exponential 
coupling function $F(\phi) = \exp(-2 \sqrt{2} k_i \phi)$.
The coupling constants $k_i$ may vary for different matter species $i$,
but for simplicity we assume that it takes the same value $k$ for all
kinds of matter.
This assumption is taken in accordance to general relativity where
gravitation couples universally to all matter species and thereby
ensures that the weak equivalence principle is satisfied.
The presence of a scalar field would lead to an additional (or {\it fifth})
force and consequently, a matter-dependent scalar coupling would lead to
violations of the weak equivalence principles.
A possible model to explain such a matter-dependent scalar coupling is given
by Damour and Donoghue \cite{damour.2010}.

In contrast to Brans-Dicke, chameleons have a scalar potential, giving the 
field a mass and therefore a finite range.  Typically, runaway potentials 
like an inverse power-law potential $V \sim \phi^{-n}$ are considered.
The interplay between such a potential and the exponential coupling causes 
the range of the scalar field to depend on the surrounding matter density.
In a dense region, like inside the Earth or within its atmosphere, the scalar 
field becomes so massive that the force corresponding to the scalar field 
becomes short ranged. This hiding feature makes it very difficult to detect 
the chameleon field with Earth-based experiments. On larger scales the field 
is long ranged and it might be detectable by experiments performed in space. 

The exterior scalar field generated by a compact object like a planet or a star
is determined only by the very outer layer of the object, we say that it has a 
thin shell. It is shown in \cite{chameleon.khoury.prl,*chameleon.khoury} that 
the exterior field is
\begin{align}
\phi(r) = \phi_0 - 3 \delta \sqrt{2} k \frac{G M_\text{E}}{r} 
e^{-m_\text{E} (r-R)},
\end{align}
where $\delta := \Delta R/R \ll 1$ is the thin shell parameter.
The chameleon field profile corresponds to the field 
\eqref{equ general scalar field profile} with $\xi = 3 \delta$, giving the 
parameter
\begin{align}
\label{equ chameleon gamma}
\gamma(r)
= \frac{ 1 - 6 \delta k^2 e^{-m_\text{E} (r-R)} }
{1 + 6 \delta k^2 e^{-m_\text{E} (r-R)} }.
\end{align}
This is the same result as found in \cite{hees}.
Furthermore, $\beta=1$ holds since the coupling is mediated by an 
exponential function.

The thin shell factor is proportional to 
$(\phi_\infty - \phi_c)/k \Phi_c$ where $\Phi_c = G M/R$ is the 
Newtonian potential of an object at its surface or, in other words, its 
compactness. $\phi_c$ and $\phi_\infty$ are the field values inside and 
infinitely far away from the compact object. They are density dependent 
and therefore the thin shell parameter depends on the composition of an object.
Typically it holds that $\phi_\infty \gg \phi_c$, such that approximately 
$\delta \sim \phi_\infty/k \Phi_c$, allowing us to compare the ability of 
testing chameleons around different compact objects in the solar system just 
by comparing their Newtonian potentials $\Phi_c$.
From this point of view, the Sun is not a promising candidate to probe 
chameleons due to its high compactness. 
The Earth, and even better the Moon, are more appropriate.

The Cassini experiment can be used to constrain the 
$(\delta_\text{Sun},m_\text{E})$-parameter space for fixed $k$ using equation 
\eqref{equ chameleon gamma}. For $k \sim 1$ and small masses for the scalar 
field, this constraints $\delta_\text{Sun}$ to the $10^{-6}$ level.
For larger masses the thin-shell factor may take much larger values.
A constraint of $\gamma$ in Earth orbit would produce the analogous result but 
for the thin-shell factor of the Earth.

It is important to keep in mind that also a satellite which aims to probe 
gravity is not a test mass and can therefore acquire a thin shell itself. 
This would further suppress any GR-violating signals.
Khoury and Weltman estimate that a typical satellite
does not have a thin shell if the condition
$10^{-15} < \delta < 10^{-7}$
is satisfied
\cite{chameleon.khoury.prl,*chameleon.khoury}.

In \cite{hees} it is argued that chameleons are ruled out due to the 
incompatibility of solar system and cosmological constraints.
But anyway, they provide an interesting example of a theory predicting 
deviations from general relativity which depend not only on the distance 
from some massive object but also on its mass, radius and composition.
It is not only important to probe gravity to high levels of accuracy,
but also around different celestial bodies.


\section{Measuring PPN parameters in Earth's exterior field}
\label{Constraining PPN Parameters in Earth Orbit}
In 2016, the Atomic Clock Ensemble in Space (ACES) mission will place an atomic
clock on the International Space Station (ISS) that is expected to reach a 
fractional frequency uncertainty of $\Delta f/f \sim 10^{-16}$ \cite{ACES}. 
In the future, space clocks will continue to improve.
After ACES, there are plans to put an optical clock on the ISS as part of the 
Space Optical Clock (SOC) project. The best optical clocks on Earth have 
already reached accuracies of $\Delta f/f \sim 10^{-18}$ over an integration 
period of $25000$ sec \cite{Poli:2014dn, Bloom:2014bc}, and significant 
progress is being made towards building optical clocks that are mobile, 
more compact and more reliable.

In this section we investigate the effect that the PPN parameters have on a 
satellite that carries an atomic clock and orbits the Earth. 
In this experiment, a precise clock on a satellite broadcasts
tick signals down to a terrestrial receiving station which records their 
arrival times using a local, more accurate clock.  
The rate at which the ticks arrive is the redshift. 
This setup allows the orbit to be tracked down to the clock accuracy. 
For given Keplerian initial conditions, we simulate both the general 
relativistic orbit as well as the orbit in an alternative theory of gravity 
with parameters different from those of general relativity. 
This solves the forward problem, and taking the difference of these 
two signals provides a way to give upper limits on how well the PPN 
parameters can be constrained by this type of mission. 
To investigate PPN parameter predictability more thoroughly, the full 
inverse problem needs to be solved, which entails reconstructing the 
full four-dimensional trajectory of the satellite by fitting different 
models to redshift data. Mock redshift data 
can be generated from solutions to the forward problem with different 
parameters and added noise. We leave attempts to solve the inverse problem 
to future work.

We choose an eccentric orbit like that originally proposed for STE-QUEST
\cite{STE-QUEST.Yellow.book,*SQ.science}. 
We solve the forward problem using the code introduced by
 Ang\'elil {\it et al.} \cite{clock.paper, angelil.saha}. Note 
that the effects that the PPN parameters have on the orbit dominate, while 
their effects on the light path between the emitter and the receiver are 
about two orders of magnitude smaller \cite{clock.paper, angelil.saha}. 
  
The trajectory of a spacecraft in Earth's external field is found by 
integrating Hamilton's equations. We have seen that for general scalar-tensor
theories the PPN parameters depend on the location where they are tested.
If the potential is set to zero, making the field massless, the PPN parameters 
$\gamma$ and $\beta$ are constant 
(see section \ref{sub:Eddington Robertson Metric}).
The corresponding metric in the Jordan frame is
\begin{subequations}
\begin{align}
&g_{tt}	= - 1 + \frac{2GM}{r}\epsilon - \frac{2 G^2 M^2}{r^2} 
\left( \beta - \gamma \right) \epsilon^2
\\
&g_{rr} = 1 + \frac{2GM}{r} \gamma \epsilon
\\
&g_{\theta\theta} = r^2
\\
&g_{\varphi\varphi} = r^2 \sin^2\theta,
\end{align}
\end{subequations}
where we consider non-isotropic Schwarzschild coordinates.
(We write $r$ instead of $\chi$ for the radial coordinate
and drop all J-indices.)
This is a special 
case of \eqref{STT metric NON-isotropic coords} with $A(r)=1,B(r)=\beta$ and 
$C(r)=\gamma$. The corresponding Hamiltonian for a satellite's trajectory in 
Earth's external field is obtained from \eqref{equ:Hamiltonian general STT}
\begin{widetext}
\begin{align}
\label{equ: Hamiltonian code}
\begin{split}
H = &- \frac{p_t^2}{2} + \left[- \frac{GM p_t^2}{r} + \frac{p_r^2}{2}
 + \frac{p_\theta^2}{2r^2}
 + \frac{p_\varphi^2}{2r^2 \sin^2\theta} \right] \epsilon
 + \left[ - \frac{2 G^2 M^2 p_t^2}{r^2} \left( 1 - \frac{1}{2}\beta
 + \frac{1}{2}\gamma \right) - \frac{G M p_r^2}{r} \gamma \right] \epsilon^2
\\
= &- \frac{p_t^2}{2} + \left[- \frac{GM p_t^2}{r}
 + \frac{\vec{p}^2}{2} \right] \epsilon
+ \left[ - \frac{2 G^2 M^2 p_t^2}{r^2}
\left( 1 - \frac{1}{2}\beta + \frac{1}{2}\gamma \right)
 - \frac{G M}{r} \frac{(\vec{x}\cdot\vec{p})^2}{r^2} \gamma \right] \epsilon^2,
\end{split}
\end{align}
\end{widetext}
where we change to Cartesian coordinates in the second line. Notice that 
$\beta$ does not show up individually, but only in combination with $\gamma$.
The equations of motion are given by Hamilton's equations.

We specify the orbit by choosing Keplerian initial conditions. We position
the Earth-clock beneath perihelion, the satellite's point of closest approach.
Hamilton's equations are integrated over 4.5 orbits, once for 
the general relativistic metric ($\gamma = \beta = 1$),
giving the redshift signal $z_\text{GR}$, and then for one where 
these parameters slightly differ from unity, giving
$z_\text{non-GR}$.
Taking the difference of the two signals, 
\begin{align}
\Delta z = z_\text{GR} - z_\text{non-GR},
\end{align}
allows us to find the maximum difference in the redshift,
$|\Delta z|_\text{max}$,
averaged over one orbit.
Such a difference in the redshift signal should be detectable if 
this residual redshift is within the accuracy of the experiment.

There are numerous both relativistic and non-relativistic
effects which enter the dynamics that have not been
considered here.
They will need to be accurately modeled as part 
of the parameter recovery procedure. 
Non-relativistic effects include
atmospheric drag, solar radiation pressure and 
Earth's Newtonian multipole field.
Ang\'elil {\it et al.} (2014) \cite{clock.paper} calculate a host of 
general relativistic effects on the satellite and the light-path trajectories. 
The $\gamma$ and $\beta$ variations discussed in this paper correspond 
to modified Schwarzschild terms in the Hamiltonian. 
The standard GR frame-dragging effect, 
the Shapiro effect (bent light paths), spin-squared 
effects on the orbit, as well as further yet weaker 
effects would need to be included when searching for deviations 
from non-GR values of $\gamma$ and $\beta$.
Effects that come into play at different orders (refer to different 
blocks in table 1 in \cite{clock.paper}) will not be degenerate with 
one another due to their fundamentally different $r-$dependence, 
provided the satellite trajectory is elliptical, inducing a sufficient 
field strength modulation over the course of the integration time.  
Further discussion on these effects may also be found in 
\cite{SQ.flyby}.

In our approach, where we subtract the 
redshift signal predicted by general relativity from that with different PPN 
parameters, all these effects will  cancel out in the subtraction process. A 
further approximation made is to allow the Earth to be transparent to the tick 
signals. In reality, however, certain portions of the experiment would miss 
data during line-of-sight loss. This would be in part compensated by having 
multiple ground stations so that at any given point a clock on Earth will be 
within the satellite's line of sight.

\begin{figure}
\centering
\includegraphics{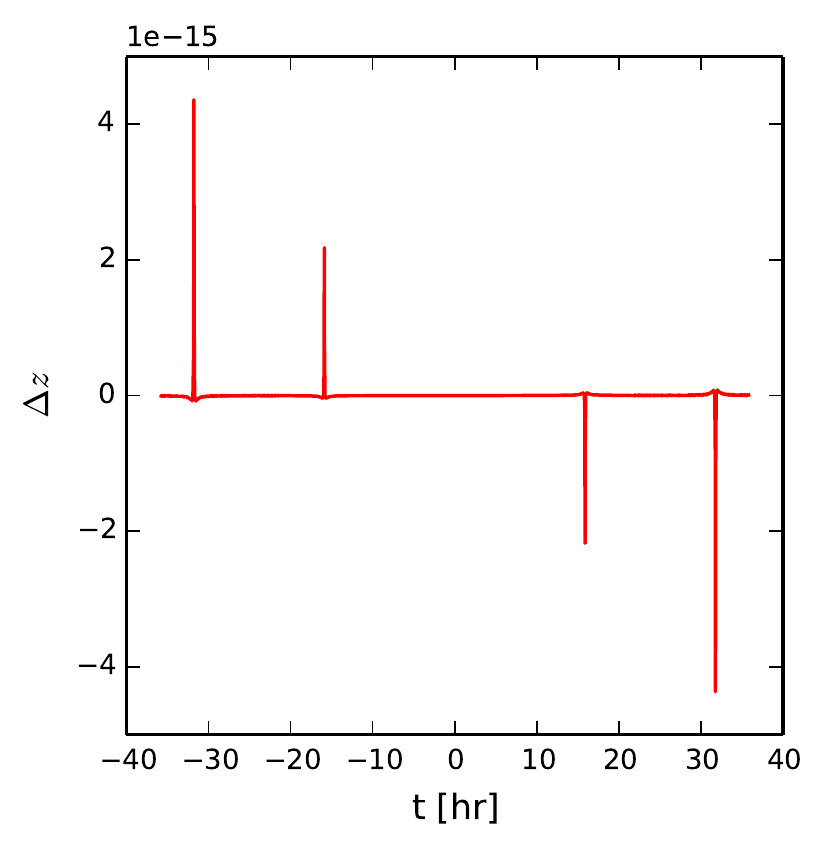}
\caption{\label{fig:Difference in Redshift Curve} Difference in Redshift Curve.
The difference in the redshift signal between the GR-orbit and the orbit with 
$\gamma=1+10^{-5}$ and $\beta=1$, $\Delta z = z_\text{GR}-z_\text{non-GR}$, 
is plotted as a function of time $t$ (in hours).}
\end{figure}

We choose an eccentric orbit with semi-major axis $a = 32'090\,\text{km}$ and
eccentricity $e = 0.779$. Such an orbit has a perihelion distance of
$7092\,\text{km}$, corresponding to an altitude of about $700\,\text{km}$ above
ground.
This orbit was chosen for the original proposal of the satellite mission
STE-QUEST \cite{STE-QUEST.Yellow.book,*SQ.science} and we take it as our
reference orbit. We then compare the general relativity orbit
to the orbit with PPN parameters differing from unity by
subtracting the redshift signal of the modified orbit from the general
relativistic orbit. Figure \ref{fig:Difference in Redshift Curve} shows the 
result for the choice $\gamma = 1+10^{-5}$, $\beta=1$.

We find that the difference peaks around pericenter, and builds up with every
orbit. For just one orbit we can read off the maximum difference in redshift 
$\Delta z = 2\cdot10^{-15}$.

\begin{figure}[h]
\begin{center}
\includegraphics{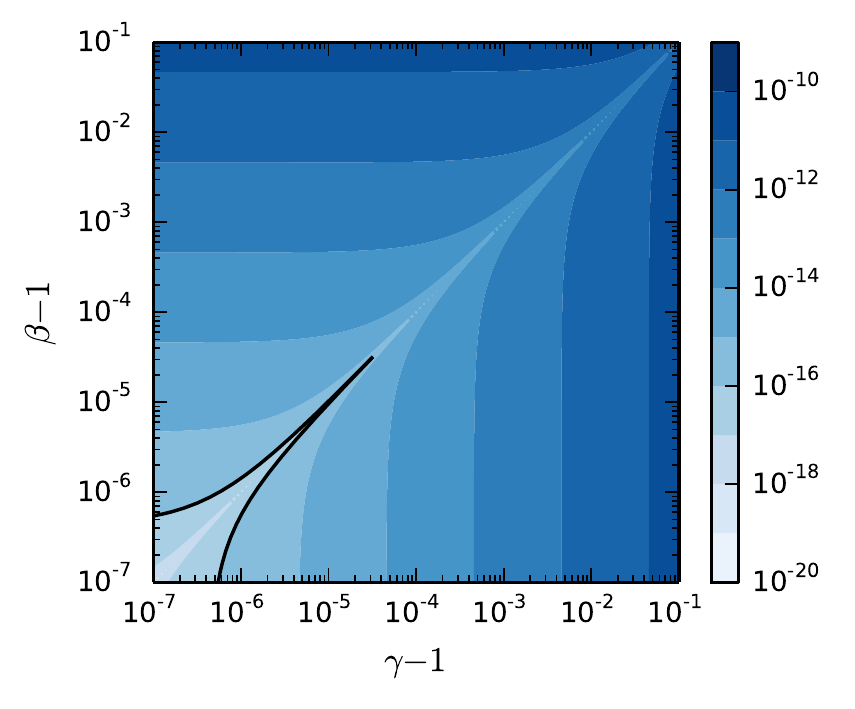}
\caption{\label{fig:log parameter plot} Logarithmic parameter space plot.
This parameter space plot shows the maximum difference in the redshift between 
the GR-orbit and that for a range of positive values for $\gamma$ and $\beta$
over one orbit.
The solid line corresponds to the value $10^{-16}$.
For the orbit we chose our reference orbit.
}\end{center}
\end{figure}

The absolute value of the maximum difference in the redshift signal over one 
orbit, indicated by its color/grey scale, is plotted for a range of parameters
in figure \ref{fig:log parameter plot}.
It is evident that, theoretically, using a clock of accuracy 
$\Delta f/f \sim 10^{-16}$ one should be able to constrain
$|\gamma - 1| \sim |\beta -1|  \sim 10^{-6}$.

Along lines with $\beta - \gamma = \text{constant}$ the absolute value of the 
signal is the same. This comes from the fact that the signal is mainly caused 
by the term in the Hamiltonian \eqref{equ: Hamiltonian code} proportional to 
$\tilde{\beta} := 1-(\beta - \gamma)/2$, while the effect of the one 
proportional to $\gamma$ is negligible. Therefore, $|\tilde{\beta}-1|$ remains 
the same if $\beta$ and $\gamma$ are interchanged, while the sign of the 
difference in the redshift signal flips.

\begin{figure*}
\begin{center}
\includegraphics{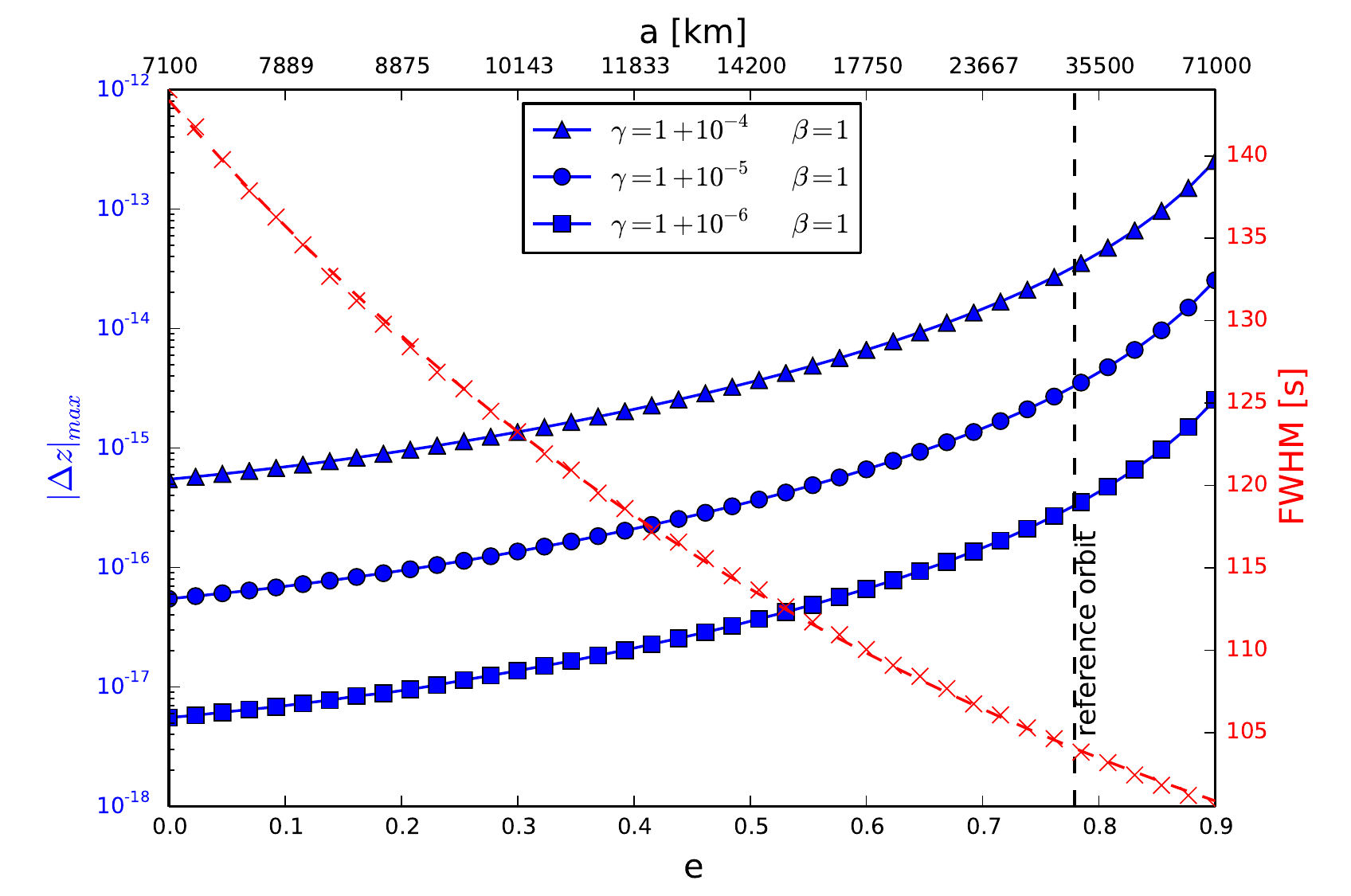}
\caption{\label{fig:SMA and e vs RS signal} 
Redshift signal and peak width as a function of eccentricity.
We compare the GR-orbit to ones where $\gamma$ slightly deviates from one.
While the pericenter distance is fixed at $d = 7100\,\text{km}$, i.e. about 
$700\,\text{km}$ above ground, the eccentricity $e$, or equivalently 
the semi-major axis $a$, is changed.
The triangle, circle and square data points show the maximum difference in
the redshift signal, $|\Delta z|_\text{max}$, for one orbit.
The cross data points 
show the full width at half maximum (FWHM) for a signal peak for 
$\gamma = 1+10^{-5}$ and $\beta = 1$. 
The analogous for other choices of the parameters are omitted 
since they would yield the same result: the width is essentially constant for 
varying PPN parameters.
We notice that the duration of the peak is of order $100$ seconds for all 
eccentricities. This is the time scale which needs to be resolved to detect 
possible variations of the PPN parameters from their GR values.
}\end{center}
\end{figure*}

Thus, having a clock on our reference orbit would allow to perform interesting 
tests of gravitational effects. It is instructive to examine several kinds of 
orbits to see which ones provide the strongest residuals. On the one hand, we 
want the satellite to pass by the Earth closely, therefore having a small 
pericenter distance in order to have strong gravitational effects. On the other
hand, it should be far enough to minimize effects as inhomogeneities of the 
Earth's gravitational potential or atmospheric drag \cite{SQ.flyby}. We fix 
the pericenter distance at $d=700\,\text{km}$ above the ground. Then, we vary
the eccentricity from a circular orbit $e=0$ to a highly eccentric orbit 
$e=0.9$, or equivalently, we vary the semi-major axis $a$ from the pericenter 
distance (circular orbit) up to $71'000\,\text{km}$ . These quantities are 
related by $d = a(1-e)$. In figure \ref{fig:SMA and e vs RS signal}, the 
maximum difference in the redshift signal over one orbit between general
relativity and some scalar-tensor theories with different 
$\gamma\neq 1$ are shown as a function of the eccentricity and the semi-major 
axis. We notice that for increasing eccentricity the 
magnitude of the signal increases significantly. 

\begin{figure}[h]
\begin{center}
\includegraphics{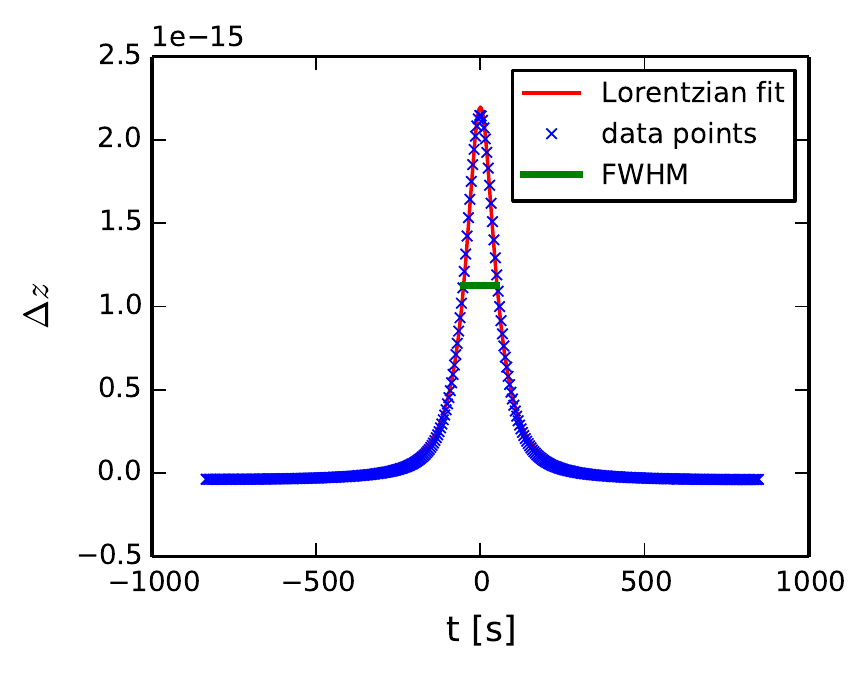}
\caption{\label{fig:Lorentzian fit} Lorentzian fit of a peak.
The data points show the difference in the redshift signal between a GR-orbit 
and $\gamma=1+10^{-5},\, \beta=1$, as a function of time (in seconds), 
centered around pericenter. 
A Lorentzian $f(t) =A/(2\pi)\Gamma[(t-t_0)^2+\Gamma^2/4]^{-1}+d$ is fitted, 
allowing to determine the full width at half maximum.
}\end{center}
\end{figure}

Now, we investigate the widths of the peaks of the difference in the 
redshift signals. The peaks are approximated by fitting a Lorentzian 
$f(t) =A/(2\pi)\Gamma[(t-t_0)^2+\Gamma^2/4]^{-1}+d$, an example is shown in 
figure \ref{fig:Lorentzian fit}.
From the fit we can easily determine the full width at half maximum.
In figure \ref{fig:SMA and e vs RS signal}, the peak width is plotted against 
the eccentricity for the case $\gamma = 1+10^{-5}$ and $\beta=1$. Even though 
the width decreases for growing eccentricities, its value is remains of order 
$\sim 100$ seconds: this is the time scale that needs to be resolved in order 
to find deviations coming from non-unity PPN parameters.
While the width depends on the orbit, it is essentially independent of the 
PPN parameters, as the values change very little in the investigated range.

\section{Conclusion}
\label{Conclusion}

We calculate the PPN parameters $\gamma$ and $\beta$ for scalar-tensor 
theories formulated in the Einstein frame for the case of a point-like source.
This extends the discussion of such theories in the Jordan frame
given in \cite{hohmann.ppn}.
To discuss tests of gravitation in the vicinity of more realistic sources
we introduce a simple formalism which can take into account
effects arising from the finite size of the source.
We use the Cassini limit on PPN $\gamma$ to put constraints on this formalism.
In particular, we update the constraints on the parameter space 
of massive Brans-Dicke scalar fields by replacing the assumption 
of a point source with that of a constant-density sphere.
This provides more stringent constraints since the proximity to the source 
is increased due to the extended radius of the object.

We emphasize that the presence of a scalar potential makes the field finitely 
ranged and therefore it is crucial to perform tests of gravitation at different
distances. 
Additionally, performing experiments around different sources
is particularly interesting because the exterior field profile is 
likely to depend on properties of an object like its compactness 
or its composition.

In the second part of the paper we discuss the possibility of testing 
scalar-tensor theories in Earth orbit using atomic clocks. Their rapid 
development and the current interest in satellite missions carrying such clocks
opens the possibility to perform comprehensive tests of gravitation within the
next decade. Such missions will provide constraints on the PPN 
parameters in the vicinity of the Earth. We calculate the relativistic effects
on the satellite orbit coming from non-GR parameters $\gamma$ and $\beta$. 
High-performance atomic clocks are sensitive to the associated change in the 
redshift signal. We find that with currently available clock technology and 
reasonable choices of spacecraft orbits one should be able to constrain 
$|\gamma - 1| \sim |\beta -1|  \sim 10^{-6}$. 
Our estimates provide upper limits to PPN parameters that could 
be measured by a clock in orbit. However, in order to provide more definite 
answers on possible constraints, one would have to solve the full inverse 
problem, where the relevant parameters are reconstructed from a redshift signal
that contains all relevant effects. We show that a PPN parameter varying from 
one produces a change in the redshift signal, peaking around pericenter of the 
eccentric orbit. While the magnitude of the peak is determined by both the 
value of the parameters and the chosen orbit, its width, and therefore the 
time-scale which needs to be resolved, depends only on the 
orbit specifications.

\begin{acknowledgments}
We acknowledge support from the Swiss National Science Foundation. 
R.B. also received support from the Dr. Tomalla Foundation.
We thank the referee for constructive comments.

\appendix

\section{Conformal transformation between Jordan and Einstein frame}
\label{app Conformal Transformation between Jordan and Einstein Frame}

In this section, we discuss the conformal transformation relating the metrics
in the Jordan and the Einstein frame. Starting in the Jordan frame 
(the converse is equivalent), we define the Einstein frame metric by 
$g_{\mu\nu}^\text{E} := \varphi g_{\mu\nu}^\text{J}$.
For the square-root of the trace of the metric and the Ricci scalars it holds
$\sqrt{-g_\text{J}} = \varphi^{-2} \sqrt{-g_\text{E}}$
and
$R_\text{J} = \varphi \left[ R_\text{E} + 6 \nabla_\text{E}^2 \ln \varphi^{1/2}
- 6 (\nabla_\text{E} \ln \varphi^{1/2} )^2 \right]$,
respectively \cite{fujii.maeda}.
Plugging this into the Jordan frame action \eqref{STT action J frame} and 
integrating by parts yields
\begin{align*}
S = &\int d^4x \sqrt{-g_\text{E}} \frac{M_\text{Pl}^2}{2} \left[ R_\text{E}
- \frac{2 \omega + 3}{2 \varphi^2} \left(\nabla_\text{E} \varphi\right)^2
- V \right]
\\
&+ \int d^{4}x \sqrt{-g_\text{E}} \mathcal{L}_\text{m}^\text{E} 
(\Phi_\text{m},\varphi^{-1} g_{\mu\nu}^\text{E}),
\end{align*}
where we defined $V := \varphi^{-2} U$ and 
$\mathcal{L}_\text{m}^\text{E} := \varphi^{-2} \mathcal{L}_\text{m}^\text{J}$.
To bring this into the desired form \eqref{STT action E frame} we define a 
new scalar field $\phi$ by demanding
\begin{align}
- 2 (\nabla_\text{E} \phi )^2 = -\frac{2 \omega + 3}{2 \varphi^2} 
\left(\nabla_\text{E} \varphi\right)^2,
\end{align}
implying
\begin{align}
\left(\frac{\partial\phi}{\partial\varphi} \right)^2
= \frac{2 \omega + 3}{4 \varphi^2}.
\end{align}
Defining the Einstein frame coupling function by $F := \varphi$, we obtain
\begin{align}
\left(\frac{\partial F}{\partial\phi} \right)^2
= \frac{4 F^2}{2 \omega + 3}.
\end{align}
This requires $\omega>-3/2$ everywhere, and therefore $\omega_0>-3/2$.
Solving for $\omega$ yields
\begin{align}
\label{equ:relation E and J frame fields}
\omega = 2 F^2 \left(\frac{\partial F}{\partial \phi} \right)^{-2} 
- \frac{3}{2}.
\end{align}
Using $F = \varphi$ and expanding both expressions in powers of $\epsilon$
\begin{subequations}
\begin{align}
F &= F_0 + F_1 (\phi - \phi_0) + F_2 (\phi - \phi_0)^2
\label{equ:E frame coupling expanded}
\\
\varphi &= \varphi_0 + \varphi_1 \epsilon + \varphi_2 \epsilon^2,
\end{align}
\end{subequations}
one obtains
\begin{align}
\begin{split}
&\varphi_0 = F_0
\qquad
\varphi_1 = F_1 \phi_1
\qquad
\varphi_2 = F_2 \phi_1^2 + F_1 \phi_2
\\
&\phi_1 = \frac{1}{F_1} \varphi_1
\qquad
\phi_2 = \frac{1}{F_1} \varphi_2 - \frac{F_2}{F_1^3} \varphi_1^2.
\end{split}
\end{align}
The relations between the coefficients of the couplings in the 
two frames are given by
\begin{align}
\label{equ coupling functions relations E J frame}
\begin{split}
&\omega_0 = \frac{2 F_0^2}{F_1^2} - \frac{3}{2}
\qquad
\omega_1 = \frac{4 F_0}{F_1^2} - \frac{8 F_0^2 F_2}{F_1^4}
\\
&F_0 = \varphi_0
\qquad
F_1 = \pm
\frac{2\varphi_0}{\sqrt{2\omega_0+3}}
\\
&F_2 = \frac{2\varphi_0}{2\omega_0+3} 
\left(1 - \frac{\varphi_0 \omega_1}{2\omega_0+3} \right),
\end{split}
\end{align}
and for the potentials, using $U = F^2 V$, one finds
\begin{align}
\label{equ potentials relations E J frame}
\begin{split}
&U_2 = \frac{F_0^2}{F_1^2} V_2
\qquad
U_3 = \frac{2 F_0}{F_1^2} \left( 1 - \frac{F_0 F_2}{F_1^2} \right) V_2 
+ \frac{F_0^2}{F_1^3} V_3
\\
&V_2 = \frac{4}{2\omega_0 + 3} U_2
\\ 
&V_3 = \pm \frac{8}{(2\omega_0+3)^{3/2}} 
\left[ -\left( 1 + \frac{\varphi_0 \omega_1}{2\omega_0 + 3}  \right) U_2 
+ \varphi_0 U_3 \right].
\end{split}
\end{align}
The coordinates in the two frames are related by
$t_\text{J} = t_\text{E}/\sqrt{F_0}$ and $\chi = r/\sqrt{F_0}$.
Note, there is a $\pm$-ambiguity when going from the Jordan to the Einstein 
frame: two theories in the Einstein frame related by 
$F_1,V_3 \leftrightarrow -F_1,-V_3$ correspond to the same theory in the 
Jordan frame.


\section{Metric in non-isotropic coordinates}

The PPN parameters are defined by introducing parameters to the individual 
terms of the expanded Schwarzschild metric written in isotropic coordinates.
But often it is useful to consider the metric expressed in non-isotropic 
coordinates. This is achieved by defining a new radial coordinate $r$ while 
the other coordinates remain the same.
(Don't confuse the notion of $r$ with the radial coordinate in the 
Einstein frame used earlier on.)
We write the metric in isotropic coordinates in the general form
\begin{subequations}
\begin{align}
g_{tt}	&= - \left(1 - \frac{2GM}{\chi} A(\chi) \epsilon 
+ \frac{2 G^2 M^2}{\chi^2} B(\chi) \epsilon^2 \right) 
+ \mathcal{O}(\epsilon^3)
\\
g_{\chi\chi} &= 1 + \frac{2GM}{\chi} C(\chi) \epsilon + \mathcal{O}(\epsilon^2)
\\
g_{\theta\theta} &= \left( 1 + \frac{2GM}{\chi} C(\chi) \epsilon \right) \chi^2
+ \mathcal{O}(\epsilon^2)
\\
g_{\varphi\varphi} &= 
\left( 1 + \frac{2GM}{\chi} C(\chi) \epsilon \right) \chi^2 \sin^2\theta  
+ \mathcal{O}(\epsilon^2).
\end{align}
\end{subequations}
By introducing a new radial coordinate
\begin{align}
r := \chi \left(1 + \frac{2 G M}{4 \chi} C(\chi) \epsilon \right)^2,
\end{align}
which can be inverted to (outside the Schwarzschild radius)
\begin{align}
\chi = r \left( \frac{1}{2} - \frac{GM}{2r} C(r) \epsilon 
+ \frac{1}{2} \sqrt{1-\frac{2GM}{r}C(r) \epsilon} \right),
\end{align}
we obtain
\begin{subequations}
\label{STT metric NON-isotropic coords}
\begin{align}
g_{tt} &= - 1 + \frac{2GM}{r} A(r) \epsilon - \frac{2 G^2 M^2}{r^2} 
\left[ B(r) - A(r) C(r) \right] \epsilon^2
\\
g_{rr} &= 1 + \frac{2GM}{r}\left[ C(r) - C'(r) r \right] \epsilon 
\\
g_{\theta\theta} &= r^2
\\
g_{\varphi\varphi} &= r^2 \sin^2\theta.
\end{align}
\end{subequations}
Here, we used that $g_{\chi\chi} d\chi^2 = g_{rr} dr^2$.
Transforming to Cartesian coordinates, the metric becomes
\begin{subequations}
\begin{align}
g_{tt} &= - 1 + \frac{2GM}{r} A(r) \epsilon - \frac{2 G^2 M^2}{r^2} 
\left[ B(r) - A(r) C(r) \right] \epsilon^2
\\
g_{x_i x_j} &= \delta_{ij} + \frac{2GM}{r}\left[ C(r) - C'(r) r \right] 
\frac{x_i x_j}{r^2} \epsilon
\end{align}
\end{subequations}
where we used
$F(r) dr^2 + r^2 d\theta^2 + r^2 \text{sin}^2\theta d\varphi^2
=  d\vec{x}^2 + \left[F(r)-1\right] \left( \vec{x}/r d\vec{x}\right)^2.$


\section{Hamiltonian}

The Hamiltonian is given by $H=1/2 g^{\mu\nu} p_\mu p_\nu$, where $p_\mu$ is 
the canonical four-momentum. Here, we consider the metric 
\eqref{STT metric NON-isotropic coords}, which we expand in powers of 
$\epsilon \sim G M/r$. The orbital velocity of a non-relativistic particle in 
a weak gravitational field is $v \approx \sqrt{G M/r} \sim \epsilon^{1/2}$, 
requiring 
$p_r, p_\theta/r, p_\varphi/(r \sin\theta) \sim v \sim \epsilon^{1/2}.$
Plugging the inverse metric into the formula for the Hamiltonian and assigning 
the terms to the appropriate orders in $\epsilon$ yields
\begin{align}
\label{equ:Hamiltonian general STT}
\begin{split}
H = &- \frac{p_t^2}{2} + \left[- \frac{GM p_t^2}{r}A(r) + \frac{p_r^2}{2} 
+ \frac{p_\theta^2}{2r^2} + \frac{p_\varphi^2}{2r^2 \sin^2\theta} \right] 
\epsilon
\\
&+ \bigg[ - \frac{2 G^2 M^2 p_t^2}{r^2} \left( A(r)^2 - \frac{1}{2}B(r) 
+ \frac{1}{2}A(r)C(r) \right)
\\
&- \frac{G M p_r^2}{r} \left( C(r) - r C'(r) \right) \bigg] \epsilon^2.
\end{split}
\end{align}
From this it is evident why we drop all terms in the spatial metric 
components that are second and higher order in $\epsilon$: they contribute 
to the Hamiltonian at third and higher orders. Notice that the expansion of 
the Hamiltonian for a signal propagating in the same spacetime looks different,
since, even though we start with the same Hamiltonian, some terms contribute at
different orders. This comes from the fact that photons travel with the speed 
of light and therefore,
$p_t, p_r, p_\theta/r$ and $p_\varphi/(r \sin\theta)$ are of order $1$.
The equations of motion are given by Hamilton's equations 
$dp_\mu/d\lambda=-\partial H/\partial x^\mu$ and 
$dx^\mu/d\lambda=\partial H/\partial p_\mu$.


\end{acknowledgments}

\bibliography{ms}

\end{document}